\documentclass[10pt]{article} 
\usepackage{amsfonts,amssymb,slashed,latexsym,setspace,cancel,subfigure}
\usepackage{graphicx,graphics,amssymb,epsf,rotate} 
\usepackage{times,cite,color,cancel}
\usepackage{hyperref}
\begin{document}

\begin{flushright}
SINP/TNP/2013/01
\end{flushright}

\begin{center}
{\Large \bf Burgeoning the Higgs mass to 125 GeV through
  messenger-matter interactions in GMSB models} \\
\vspace*{1cm} \renewcommand{\thefootnote}{\fnsymbol{footnote}} { {\sf
    Pritibhajan Byakti${}^{1}$} and {\sf Tirtha Sankar Ray${}^{2}$}}
\\
\vspace{10pt} {\small ${}^{1)}$ {\em Saha Institute of Nuclear
    Physics, 1/AF Bidhan Nagar, Kolkata 700064, India} \\ 
${}^{2)}$ {\em ARC Centre of Excellence for Particle Physics at the Terascale,
School of Physics, \\University of Melbourne, 
Victoria 3010, Australia}} 
\normalsize
\end{center}

%%%%%%%%%%%%%%%%%%%%%%%%
\begin{abstract}
 A 125 GeV Higgs renders the simpler GMSB models unnatural,
 essentially pushing the soft spectrum beyond the LHC reach. A direct
 coupling of the matter and messenger fields, that facilitates an
 enhanced mixing in the squark sector, is a way to ameliorate this
 deficiency.  We construct all possible messenger-matter interaction
 terms considering the messenger multiplets in 1, 5 and 10 dimensional
 representations of the SU(5). A Froggatt-Nielsen like flavor
 framework connected with the origin of fermion mass hierarchy is
 utilized to control the interaction terms and suppress FCNC.  We
 perform a detailed comparative study of the efficiency of such
 interaction terms to boost the Higgs mass keeping the soft spectrum
 light. We identify the more promising models and comment on their
 status in present and future collider studies.
\end{abstract}

\setcounter{footnote}{0}
\renewcommand{\thefootnote}{\arabic{footnote}}

%%%%%%%%%%%%%%%%%%%%%%%%
\section{Introduction:}~
Identification of the recently observed \cite{:2012gk,:2012gu} scalar
field at the LHC with the Higgs would cast a long shadow on Gauge Mediated
Supersymmetry Breaking (GMSB) models
\cite{Draper:2011aa,Giudice:1998bp}. This is a direct consequence of
the well known phenomenon that in pure gauge mediation models, mixing
in the scalar sector is minimum. This implies that in order to raise
the Higgs mass to $\sim 125$ GeV one needs stop masses at several TeV,
castigating these models to an \textit{unnatural} existence with bleak
chances to be probed at collider experiments like the LHC.

A resolution of this predicament is to consider direct coupling
between the messenger and matter fields or models with gauge
messengers \cite{Intriligator:2010be}.  These scenarios can in
principle generate a sizable trilinear coupling that can boost the
Higgs mass given by,
\begin{eqnarray}
\label{mh}
 m_h^2 = M_Z^2 \cos^2 2\beta +
 \frac{3}{4\pi^2}\frac{m_t^4}{v^2}\left[\log\frac{M_S^2}{m_t^2} +
   \frac{X_t^2}{M_S^2} \left(1-\frac{X_t^2}{12M_S^2}\right)\right] \,
 ,
\end{eqnarray}
where, $X_t=A_t - \mu \cot\beta $ and
$M_s=\sqrt{m_{\tilde{t}_1}m_{\tilde{t}_2}}$ , while keeping the scalar
spectrum within the range of interest for collider physics. In case of
gauge messengers the  trilinear couplings are proportional to gauge
charges. However realistic models recently studied in
\cite{Bajc:2012sm} predict a relatively heavy spectrum. In this paper
we will consider models of direct messenger-matter interactions that
can lead to relatively large trilinear coupling and a considerably
light soft spectrum making them  more pleasing from a
fine-tuning point of view and more interesting phenomenologically.
These interaction terms generally lead to
new contributions to the scalar masses, thus producing a correlated
perturbation of the pure gauge mediation soft spectrum.  The strength of the
interaction and hence the size of the trilinear coupling is
principally constrained from the considerations to keep the scalar masses
non-tachyonic and achieve radiative electroweak symmetry breaking (REWSB).
Also a cause for concern are the
complications in the flavor sector
\cite{Barbieri:1995tw,Barbieri:1995rs,ArkaniHamed:1995fs}. These tend
to push the messenger scale upward into regions where the Gravitino
mass goes beyond $\sim $ sub-KeV  range putting unfavorable upper bounds on
the reheating temperature inviting strong constraints from BBN
\cite{Khlopov:1984pf}.  The flavor constraints are severe for the
first two generations of fermions and can be minimized by considering
that the messengers preferentially couple to the third generation,
which is relevant for enhancing the Higgs mass. This can be ensured by
imposing judiciously chosen flavor symmetries.

Several models of GMSB augmented with messenger-matter interactions to
alleviate the problem with the heavy Higgs have been suggested in the
literature
\cite{Kang:2012ra,Albaid:2012qk,Han:1998xy,Craig:2012xp,Abdullah:2012tq
  ,Evans:2011bea ,Martin:2012dg}.  The basic idea being the coupling of
the messengers to one of the $Q, U, H_u$ MSSM multiplets, generating a sizable
$A_t$ at the messenger scale.  Possible interaction terms in a given scenario
get determined by the content and quantum numbers of the messenger
sector. In this paper we make a systematic study of messenger-matter
interactions, possible within a well defined framework. As an
organizing strategy we  consider  the messengers as vector
pairs embedded in one of the simpler representations of the GUT group
SU(5). We will restrict ourselves to messengers in the 1, 5 and 10 dimensional
representations of SU(5) or an admixture. This ensures that the
perturbative unification \footnote{Note that it is possible to ensure perturbative gauge
   coupling unification without considering complete representations
   of the GUT groups\cite{Calibbi:2009cp}. These {\sl magic}
   combinations can potentially lead to interesting phenomenological
   scenarios\cite{Byakti:2012qk}, we will not discuss them in this
   paper.} of gauge coupling is not ruined due to
introduction of these additional fields \cite{Picariello:1998dy}.
 For the first time  we make  a quantitative comparison of various models,
 including some that have been discussed in the literature and some
 entirely new scenarios, in term of their effectiveness to raise the
 Higgs mass without the usual pitfall of large scalar masses.
 In order to make the
 comparison, we numerically scan  all parameters of every given
 model over a suitable range  and project the allowed regions on a common parameter
 space. This allows us to make precise statements about scenarios
 preferred by the recent data on the \textit{supposed} Higgs mass.

For simplicity, in our study we will not attempt to model the hidden
sector and simply assume supersymmetry is broken by the vev of a
spurion field which couples to the messengers. The crucial point would
be that the messengers other than having  usual gauge couplings to
the MSSM sector, now also couple directly through the superpotential.
In this paper, we
enumerate possible messenger-matter interaction terms allowed by a
given messenger sector. For each scenario we compute the contributions
of these new terms to the scalar masses at one and two loop order. We
find that the one loop contributions are always tachyonic but they are
suppressed by $x^2 \equiv (F/M^2)^2$ where $M$ and $F$ are messenger
scale and supersymmetry breaking scale respectively.  Thus for a given
$\Lambda=F/M$ or soft scalar mass, this contribution decouples as the
scale of supersymmetry breaking is increased. The sign of the two loop
contribution is model dependent, however there is no suppression from
the supersymmetry breaking scale. In most regions of the parameter
space this becomes the dominant contribution from the new terms. One
loop renormalization group equations are used to run these soft
parameters down from the scale of supersymmetry breaking to the weak
scale and the sparticle spectrum is generated.

The rest of the paper is organized as follows. In Section 2 we
describe our organizing principle and enumerate the possible
interaction terms. In Section 3 we compute the soft masses and
trilinear couplings for the different models. In Section 4 we describe
our numerical procedure and present our results comparing the
models. Finally we conclude.

%%%%%%%%%%%%%%%%%%%%%%%%%%
\section{Messenger-Matter interactions}
 In this section, we collect the possible interactions between the
 messenger sector and the MSSM fields. A useful way to organize these
 is to consider the fields embedded in a representation of some GUT
 group like SU(5).  As usual, the MSSM fields can be embedded into
 representations of the SU(5) group as follows,
 \begin{eqnarray*}
 \bar5&=& (\bar3,1,\frac13)\oplus(1,2,-\frac12)=D^c\oplus L,~
10=(3,2,\frac16)\oplus(\bar3,1,-\frac23)\oplus(1,1,1)=Q\oplus U^c\oplus E^c\\
\bar5_H&=&(\mbox{integrated out field})\oplus H_d,~
5_H=(\mbox{integrated out field})\oplus H_u
\end{eqnarray*}

The messenger fields are in a vector like pair embedded in 1, 5, 10 and
their conjugate representations of SU(5).  For the rest of this paper
we use the following nomenclature for the messenger sector,
\begin{eqnarray*}
 1_m=&S_m,~~~ 5_m= \tilde D_m^c\oplus H_u^m,~
\bar{5}_m=D_m^c\oplus H_d^m\\
10_m=&Q_m\oplus U^c_m\oplus E^c_m, ~\bar{10}_m =\tilde Q_m\oplus \tilde
U_m^c \oplus \tilde E_m^c,
\end{eqnarray*}
where the subscript $m$ identifies the messenger fields.

Technically the singlet is not a valid gauge messenger in the usual
sense.  However, it can couple directly to the visible sector through
superpotential terms and thus will be considered in the following
discussion.  Within this framework for the messenger and the matter
sectors, it is simple to write down all the possible SU(5) invariants
that can be constructed:
\begin{eqnarray*}
\mbox{{\bf Only Singlets}}& (i)& 5_H\bar5_H 1_m~~(ii)~5_H\bar5~ 1_m
\\ \mbox{{\bf Only $\mathbf{5\oplus\bar5}$}}& (i)&
10~\bar5_H\bar5_m~~(ii)~ 10~\bar5~\bar5_m ~~(iii)~ 10~\bar5_m\bar5_m
~~(iv)~ 10~\bar5_{mi}\bar5_{mj} ~~(v)~ 10\, 10~ 5_m\\ \mbox{{\bf Only
    $\mathbf{10\oplus \overline{10}}$}}& (i)& 10\, 10_m 5_H
~~(ii)~10_m\, 10_m 5_H ~~(iii)~\overline{10}_m \, \overline{10}_m
\bar5_H ~~(iv)~\overline{10}_m \,\overline{10}_m \bar5.
\end{eqnarray*}
Note that the invariants $\overline{10}_m \, \overline{10}_m \bar5_H$
and $\overline{10}_m \, \overline{10}_m \bar5$ will not give $A_t$
hence will not be considered further. The possibility that the
messenger sector is composed of fields in different representation can
also exist. In this case we expect the usual invariants listed above
should reappear. The new possible invariants are given below,
\begin{eqnarray*}
 \mbox{{\bf Singlet + $\mathbf{5\oplus\bar5}$}} &(i)&~5_H\bar5_m
 1_m~~(ii)~5_m \bar5 1_m~~(iii)~5_m\bar5_H 1_m \\ \mbox{{\bf Singlet +
     $\mathbf{10\oplus \overline{10}}$}} &(i)& 10\,\overline{10} 1_m
 \\ \mbox{$\mathbf{5\oplus\bar5 + 10\oplus \overline{10}}$} &(i)&
 10\,10_m\, 5_m ~~(ii)~\overline{10}_m\,5_m 5_H
\end{eqnarray*}
Again we find that terms like $5_m \bar5 S$ and $5_m\bar5_H S$ will
not give $A_t.$ To keep the discussion tractable we will utilize
prudently chosen flavor symmetries to suppress interaction terms other
than the ones listed above.

As emphasized earlier a host of issues with flavor including FCNC can
be controlled if we consider scenarios where the messenger sector
preferentially couples to the third generation multiplet. An
economical proposition is to connect this to the flavor symmetry that
is responsible for the hierarchy in measured mass of the Standard
Model fermions\cite{Albaid:2012qk}. For example one can consider the
Froggatt-Nielsen \cite{Froggatt:1978nt} framework that necessitates an
$U(1)_F$ flavor symmetry group, under which the MSSM chiral multiplets
may be charged. This flavor symmetry is spontaneously broken at some
high scale $M_{string}$ by the vev of the flavon field $(Z)$, with a
conventionally assigned flavor charge $-1$.
  For this choice,  operators with a negative $U(1)_F$ charge
  mismatch ($\Delta F$) are considered absent in the
  effective theory. However, if $\Delta F>0$ for any superpotential term
  then it is suppressed by the usual Froggatt-Nielsen 
  factor $\epsilon^{\Delta F}$ where
  $\epsilon = \langle Z \rangle /M_{string}$. Flavor symmetries of this kind
can successfully explain  observed fermion mass matrices within present experimental
uncertainties. Interestingly the
anomalous nature of the symmetry can motivate compensatory exotic
particles that can be probed at present and future collider
experiments \cite{Savoy:2010sj,Eboli:2011hr,Ray:2011dm}, thus providing an
handle to probe the  existence of these flavor structures.  Together with 
R-parity assignments this will be enough to determine the interaction terms
uniquely for most of the scenarios that will be explored in this paper. The
usual charges for the MSSM and the flavon multiplets are given in
Table~\ref{table1}.

Another symmetry that is usually useful in safely segregating the
messenger and visible sectors in usual GMSB models is the messenger
parity \cite{Dimopoulos:1996ig} under which all the messenger
multiplets are assumed odd while the MSSM multiplets are considered
even. One can classify the interaction terms introduced above into two
categories with different implications for possible messenger
parities. In models where the interaction term involves
two messenger multiplets and one MSSM multiplet one can impose
messenger parity consistently. This immediately forbids mass mixing
terms between the messenger and matter multiplets at all orders of the
perturbation theory. While for models where the interaction term
involves one messenger multiplet and two MSSM multiplets, a consistent
messenger parity cannot be constructed. This will lead to messenger-matter
mixing either at tree level or at higher orders, leading to 
non-trivial contributions to the soft spectrum. We will discuss the
consequences of this for specific scenarios in the next section.

 \begin{table}
 \begin{center}
\begin{tabular}{|c|c|c|c|c|c|c|c|}
  \hline Multiplets\vphantom{$\frac{\frac12}{\frac12}$} & $10_1$ &
  $10_2$ & $10_3$ & $\bar 5_1$ & $\bar 5_{2,3}$ & $5_H, \bar{5}_H$ & Z
  \\ \hline $U(1)_F$ \vphantom{$\frac{\frac12}{\frac12}$}& 4 & 2& 0&
  $p+1$ & p & 0 & -1\\ \hline Rp \vphantom{$\frac{\frac12}{\frac12}$}
  &\multicolumn{3}{|c|}{-1}& \multicolumn{2}{|c|}{-1} & 1 & 1
  \\ \hline
\end{tabular} 
\end{center}
\caption{ \em \small U(1) flavor symmetry and the R-parity charges of
  different MSSM multiplets.  The subscripts denote the flavor index.
  Values $p=0,1, 2$ approximately explain the fermion mass hierarchy.
  The charge assignment for the MSSM multiplets will remain same
  throughout the paper.}
\label{table1}
\end{table}

%%%%%%%%%%%%%%%%%%%%%%%%%
\section{The soft breaking masses} \label{models}
We assume that supersymmetry is broken due to some hidden sector
dynamics that can be parametrized into a spurion $(X)$ vev. The
messengers which are charges under the MSSM gauge groups, couple to
these spurion fields and communicate supersymmetry breaking to the
visible sector through the usual gauge couplings. Usually the
messengers are assumed to have some messenger parity that prevents
these fields from mixing or interacting with the visible sector. In
this paper we will relax this and consider all possible interaction
terms between the visible and hidden sectors. The entire superpotential
can be schematically written as,
\begin{equation}
 W= W_{\rm MSSM} (\phi_{vis}) + \lambda_{\phi} X\phi_m\tilde \phi_m +
 W_{int} (\phi_m,\phi_{vis}),
 \label{master_lang}
\end{equation}
where the spurion gets a supersymmetry breaking vev $\langle X \rangle
= M +\theta^2 F$ and $\{\phi_m, \tilde\phi_m\}$ are vector like pair
of the messenger fields and $\phi_{vis}$ are the usual MSSM chiral
supermultiplets. We define the messenger scale $M_{mess} =
\lambda_{\phi} M$ and $\Lambda = F/M$. The loop integrals can be
expressed in terms of the dimensionless parameter
$x=\Lambda/M_{mess}.$

The most general messenger sector however can involve mass terms for
the messenger fields of the form $m_{ij}\phi_i \tilde\phi_j.$ This can
introduce a new dimensionful parameter other than the messenger
scale. The ensuing complication in determination of the soft spectrum
through the wave-function renormalization technique lead to the so
called $m_{H_u} - A_u$ problem \cite{Craig:2012xp} in models where $
H_u$ and/or $H_d$ couple with messengers.  In the models discussed in
this paper we will find the these mass terms are either absent or
suppressed by a factor $\epsilon^{a}$, where $a$ is the flavor charge
of the spurion field $X$. We can in principle make $a$ large to
suppress these terms effectively. We will neglect these terms in our
calculations below.

The soft masses get the usual contributions from the gauge
interactions of the messenger fields at one loop for the gaugino
masses and two loops for the scalar masses, given
by\cite{Giudice:1998bp},
\begin{eqnarray}
M_r &=& d \frac{\alpha_r}{4 \pi} g(x)~ \Lambda, \nonumber
\\ M_{\phi_i}^2 &=& 2 d\sum_{r=1,2,3}\left[\left(\frac{\alpha_r}{4
    \pi}\right)^2C_r(i) \right] f(x)~ \Lambda^2,
\label{usual}
\end{eqnarray}
where $d$ is the Dynkin index and is $1$ for messengers in the
$5\oplus \bar5$ and $3$ for messengers in $10\oplus\overline{10}.$ The
$C_r(i)$'s are the usual Casimir invariants for the representation
$i$ and,
\begin{eqnarray}
f(x) &=& \left. \frac{1+x}{x^2} \right.  \left[ \log (1+x) - 2
  \mbox{Li}_2 (x/[1+x]) + \frac{1}{2} \mbox{Li}_2 (2x / [1+x]) \right]
+ (x \rightarrow -x) \nonumber, \\ g(x) &=&\left. \frac{1}{x^2} \right.
\left[ (1+x) \log (1+x) + (1-x) \log (1-x) \right].
\end{eqnarray}
Note that two loop contributions to the gaugino masses were computed 
in \cite{Picariello:1998dy,Lee:2011dw}. For messenger scale beyond 100 TeV the corrections are at 
a few percent level. We will neglect this small correction in our numerical
calculations.

Now we turn to the contribution of the messenger-matter interaction term in
Eq.~\ref{master_lang}. In this paper we will only consider interaction
terms of the form $W_{int}=\lambda_{ijk}\phi_i\phi_j\phi_k,$ where at
least one chiral multiplet from both the messenger and the visible
sector are present. The contribution at one loop level to a field
$\phi_i$ belonging to the MSSM can be directly computed and is given
by,
\begin{eqnarray}
 \left. \delta M_i^2 \right|_{1-loop} &=& -C_{ijk}
 \frac{|\lambda_{ijk} |^2}{96 \pi^2} x^2 \Lambda^2 h(x),
\label{one_loop}
\end{eqnarray}
where $C_{ijk}$ is the multiplicity factor that measures the effective
number of messenger fields that the MSSM supermultiplet $\phi_i$
couples to through the superpotential coupling $\lambda_{ijk}$
including the appropriate group theoretic factors\footnote{For
  instance, in the interaction $\lambda Q U^c_m H_U$, $C_{ijk}$
  for $Q$ is $1$ whereas $C_{ijk}$ for $H_U$ is 3 (color factor). In
  contrast for $\lambda Q_m U^c_m H_U$, $C_{ijk}$ for $H_U$ is now
  $6$.}. The expression of $h(x)$ is as follows
\cite{Evans:2011bea,Craig:2012xp},
\begin{equation}
 h(x)=3\frac{(x-2)\log(1-x)-(x+2)\log(1+x)}{x^4}.
\end{equation}
Let us make the following observations regarding the one loop
contribution: (i) note that $h(x\rightarrow 0)\rightarrow1$ and thus
(ii) the one loop contribution decouples with the messenger scale
$(\left. \delta M_i^2 (x\rightarrow 0)\right|_{1-loop} \rightarrow
0)$, (iii) the contribution is always negative.

It is easier to use the wave function renormalization techniques
\cite{Giudice:1997ni,Chacko:2001km} in order to compute the the two loop 
contributions. They can be written in terms of the anomalous dimension
and the $\beta$ functions for the Yukawa coupling, below and above the
messenger scales. Adaptations of the generic framework for messenger-matter 
interaction terms were computed in
\cite{Chacko:2001km,Kang:2012ra}, we quote them for the sake of
completeness,
 \begin{eqnarray}\label{two_loop}
\left. \delta M_i^2 \right|_{2-loop} &=&\frac12
\sum_\lambda\left[\beta_{\lambda}^+\frac{\partial (\Delta
    \gamma_i)}{\partial \lambda}-\Delta\beta_{\lambda}\frac{\partial
    (\gamma_i^-)}{\partial \lambda} \right]_{M_{mess}}
\Lambda^2,\\ \left. A_{ijk}\right|_{1-loop} &=& -\left(\Delta\gamma_i+
\Delta\gamma_j +
\Delta\gamma_k \right)~h_A(x)~\Lambda,
 \end{eqnarray}
where, $\Delta X|_{M_{mess}}= [X^+-X^-]|_{M_{mess}}
=[X({M_{mess}}+\delta)-X({M_{mess}}-\delta)]|_{\delta\to0}$,
$\beta_\lambda=d\lambda/dt$ and
\begin{equation}
 h_A(x)= \frac{1}{2x}\log\left(\frac{1+x}{1-x}\right).
\end{equation}
As has been pointed out recently in \cite{Evans:2013kxa} one should be careful
to interpret the above formula for models where a consistent messenger parity
cannot be imposed and thus leads to the possibility of mixing between the
messenger and MSSM multiplets.  These kinetic mixing terms can
be removed by  an unitary rotation. The above formula gives
correct results for this case \cite{Evans:2012hg} assuming a non-standard definition
of the corresponding anomalous dimensions.
We have checked that in all the cases where such mixing can arise, our results
are in agreement with the treatment prescribed in \cite{Evans:2013kxa}.

We will now compute the mass spectrum of each of the models using the
generic expressions for the soft masses introduced in this section. We
will only indicate the new contributions arising from the interaction
terms in addition to the usual contributions given in
Eq.~\ref{usual}. For each model we will indicate the R-parity and the
flavor charges for the messenger sector. The corresponding charges for
the MSSM multiplets are given in Table~\ref{table1}. We will assume
that wherever the interaction terms include two multiplets, a messenger
parity is imposed under which the messenger sector is odd while the
multiplets of the visible sector are even.

\subsection{Only Singlets}
In these scenarios one has to assume that the messenger sector in
addition to the singlets,  also has the usual messenger fields
charged under the MSSM gauge group. As a definite choice we will
assume that along with the singlet the messenger sector consists of a
single $5\oplus \bar5$ vector pair of chiral messengers
$(\phi_m~\tilde\phi_m)$. However a messenger parity prevents them from
directly coupling with the visible sector.  These spectator messengers
will contribute to the soft masses through usual gauge interactions
according to Eq.~\ref{usual}.

\subsubsection{Model 1} $\mathbf{5_H\bar{5}_H 1_m:}$  
Considering that the singlet ($1_m$) is even under R-parity, one
obtains the following messenger-matter mixing superpotential term
\cite{Kang:2012ra},
\begin{equation}
 W_{int}=\lambda H_u H_d S_m.
\end{equation}
The new contributions to the soft masses can be read off from
Eqs.~\ref{one_loop} and \ref{two_loop}. They are given by,
\begin{eqnarray}
\delta M^2_Q &=& - \frac{ \alpha_\lambda}{16 \pi^2}(\alpha_t+\alpha_b)
\Lambda^2,~~ \delta M^2_{U^c} = - \frac{\alpha_t
  \alpha_\lambda}{8\pi^2}\Lambda^2, ~~\delta M^2_{D^c} = -
\frac{\alpha_b\alpha_\lambda}{8 \pi^2} \Lambda^2, \nonumber \\ \delta
M^2_L &=& - \frac{\alpha _{\lambda } \alpha _{\tau }}{16 \pi
  ^2}\Lambda^2,~~ \delta M^2_{E^c} = -
\frac{\alpha_\lambda\alpha_\tau}{8 \pi^2} \Lambda^2, \nonumber
\\ \delta M^2_{H_u} &=& \left[ -\frac{\alpha_\lambda}{24\pi}x_1^2
  h(x_1)+\frac{\alpha_\lambda}{16 \pi^2} \left( 4 \alpha_\lambda +
  \alpha_\tau + 3\alpha_b -3\alpha_2 -\frac35
  \alpha_1\right)\right]\Lambda^2,\\ \delta M^2_{H_d} &=& \left[
  -\frac{\alpha_\lambda}{24\pi}x_1^2 h(x_1)+ \frac{\alpha_\lambda}{16
    \pi^2} \left( 4 \alpha_\lambda +3 \alpha_t -3\alpha_2 -\frac35
  \alpha_1\right)\right]\Lambda^2, \nonumber \\ A_t &=& A_b=A_\tau= -
h_A(x)\frac{\alpha_\lambda}{4 \pi}\Lambda, \nonumber
\end{eqnarray}
where $\alpha_{\lambda_i} = \lambda_i^2/4 \pi$ and the subscripts have
their usual meaning. We will follow this convention through out the
paper.  Unfortunately as commented in \cite{Kang:2012ra}, one can
anticipate an anomalous contribution to the $\mu-B_{\mu}$ parameters 
in this model.
%%%%%%%%% 
\subsubsection{Model 2}
%%%%%%%% 
$\mathbf{5_H \bar5 1_m:}$ The other possibility here is to assume that
the messenger field is odd under the R-parity. However in this case we
get contributions to the neutrino mass arising through type I see-saw
mechanism\footnote{See \cite{Joaquim:2006uz,Joaquim:2006mn}
for realistic models of neutrino mass within the GMSB framework that
utilize messenger-matter interactions.}\cite{Mohapatra:1998rq}. This puts a lower bound on the
messenger scale at $\sim 10^{10} GeV$. A way to evade this to is
consider another singlet field $\tilde S$ and impose non-trivial
flavor charges:
\begin{equation}
 U(1)_F(1_m, \tilde{1}_m, 5_m, \bar{5}_m, X)=(-p,-a+p,-(a+q),q,a).
\end{equation}
We obtain the following
superpotential through which the singlet messenger field interact with the MSSM multiplets,
\begin{equation}
 W_{int}=\lambda H_u L_3 S_m.
\end{equation}
We made a field redefinition so that only the third family can
interact with the messenger field. We considered $a$ and $q$ to be
large positive numbers so that all the non-renormalizable terms in the
messenger-matter mixing sector are highly suppressed. As is clear from
the above expression, the field $S_m$ must carry a unit lepton
number. The new contributions to the soft masses from the given
interaction term is given by,
\begin{eqnarray}
\delta M^2_Q &=& - \frac{\alpha_\lambda\alpha_t}{16\pi^2}\Lambda^2,~~
\delta M^2_{U^c} = - \frac{\alpha_\lambda\alpha_t}{8\pi^2}\Lambda^2,~~
\delta M^2_{E^c} = - \frac{\alpha_\lambda\alpha_\tau}{8\pi^2}\Lambda^2,~~
\delta M^2_{H_d} = -
\frac{\alpha_\lambda\alpha_\tau}{16\pi^2}\Lambda^2, \nonumber \\ \delta
M^2_{L} &=& \left[ -\frac{\alpha_\lambda}{24\pi}x^2
  h(x)+\frac{\alpha_\lambda}{16\pi^2} (4 \alpha_\lambda +3
  \alpha_t-3\alpha_2-\frac35\alpha_1)\right]\Lambda^2,
\nonumber\\ \delta M^2_{H_u} &=& \left[
  -\frac{\alpha_\lambda}{24\pi}x^2 h(x)+\frac{\alpha_\lambda}{16\pi^2}
  (4 \alpha_\lambda
  +\alpha_\tau-3\alpha_2-\frac35\alpha_1)\right]\Lambda^2,
\nonumber\\ A_t&=&A_\tau=- h_A(x)\frac{\alpha_\lambda}{4\pi}\Lambda.
\end{eqnarray}

\subsection{Only $5\oplus\bar{5}$}
In this section we look at models where the messengers are in the
$5\oplus\bar{5}$ representations. Depending on the choice of
symmetries various invariants can be constructed. We will now study
the possible terms in turn and compute the new contributions to the
scalar masses. Again the usual contributions are given by
Eq.~\ref{usual} where we set the Dynkin index $d=1.$ Unless mentioned
we will assume that the number of generations of messenger is one.
%%%%%%%%%% 
\subsubsection{Model 3}
$\mathbf{10\bar5_H\bar5_m}:$ Let us consider that the messengers are
odd under the R-parity and the following flavor charges are imposed,
\begin{equation}
 U(1)_F(5_m,\bar{5}_m,X)=(-a,0,a).
\end{equation}
We obtain the following  interaction term in the superpotential,
\begin{equation}
 W_{int} = \lambda_q Q_3D^c_m H_d + \lambda_e E^c_3 H_d^m H_d.
\end{equation}
The absence of messenger parity allows the operator $\bar5_H\,5_m \,
X$ to be consistent with all other symmetries imposed. This can be
absorbed in the superpotential above by a basis change. However there
is still a loop level mixing between $\bar{5}_H$ and $\bar{5}_m$. This
can be rotated away at the one loop order but contributes non
trivially to two loop corrections of the soft spectrum.  The
contributions of these mixing terms to the trilinear coupling are numerically
negligible compared to the unsuppressed one loop contributions and
thus can be neglected.  We extend the analysis in
\cite{Albaid:2012qk,Han:1998xy} by including all the two loop contributions to
the scalar masses and considering the effect of one loop mixing between the
messenger and matter multiplets. The expressions are given by,
\begin{eqnarray}
\delta M^2_Q &=& \left[ -\frac{\alpha_q}{24\pi}x^2
  h(x)+\frac{\alpha_q}{16\pi^2}\left[ 6\alpha_q + 4 \alpha_b +
    \alpha_\tau + \alpha_e -\frac{16}{3}\alpha_3 -3\alpha_2-
    \frac{7}{15}\alpha_1\right] -\frac{\alpha_b \alpha_e}{16\pi^2}
  \right] \Lambda^2, \nonumber \\ \delta M^2_{U^c} &=& -
\frac{\alpha_q \alpha_t}{8 \pi ^2} \Lambda^2, ~~ \delta M^2_{D^c} = -
\frac{\alpha_b }{8 \pi^2}(\alpha_e+ 4 \alpha_q)\Lambda^2,~~ \delta
M^2_L = - \frac{3 \alpha_\tau}{16\pi^2} (\alpha_e + \alpha_q)
\Lambda^2, \nonumber\\ 
\delta M^2_{E^c} &=& \left[-\frac{\alpha_e}{12\pi}x^2
h(x)+\frac{\alpha_e}{16\pi^2}\left[ 8
    \alpha_e +6 \alpha_q + 6\alpha_b + 4\alpha_\tau -6\alpha_2
- \frac{18}{5}\alpha_1
    \right]- \frac{3\alpha_q }{8\pi^2}\alpha_\tau \right]\Lambda^2,
\nonumber \\ 
\delta M^2_{H_d} &=& \left[   -\frac{\alpha_e+3\alpha_q}{24\pi}x^2
  h(x)+\frac{\alpha_e}{16\pi^2}\left[ 4 \alpha_e +
2\alpha_\tau  -3 \alpha_2 -
    \frac{9}{5}\alpha_1 \right]\right.\nonumber\\
 &&\left. +
  \frac{\alpha_q}{16\pi^2}\left[18 \alpha_q +3 \alpha_t +12
\alpha_b      +6 \alpha_e -16\alpha_3 -9\alpha_2 -\frac75 \alpha_1 \right]
  \right]\Lambda^2,\\
 \delta M^2_{H_u} &=& - \frac{3\alpha_q
  \alpha_t}{16\pi^2} \Lambda^2, ~~A_t = - \frac{\alpha _q}{4 \pi
}\Lambda, ~~ A_b = - \frac{\alpha _e+4 \alpha _q}{4 \pi }\Lambda, ~~
A_\tau = - \frac{3 \left(\alpha _e+\alpha _q\right)}{4 \pi
}\Lambda. \nonumber
\end{eqnarray}

\subsubsection{Model 4}$\mathbf{10 10 5_m}:$ If we consider the messengers 
$(5_m~\mbox{and}~ \bar{5}_m)$ are even under the R-parity we can have
two different invariants $10\, 10\, 5_m$ and $10\, \bar 5\, \bar{5}_m$
that are possible\cite{Han:1998xy}. However depending on the assignment of the
flavor charges, one or the other might may become more dominant. We
will consider by turn the two extreme scenarios where only one of the
invariants dominates. Considering the flavor charges,
\begin{equation}
U(1)_F(5_m,\bar{5}_m,X)=(0,a,-a),
\end{equation}
the dominating part of the superpotential is given by,
\begin{equation}
 W_{int}=\lambda_q Q_3U^c_3 H_u^m+ \lambda_u U^c_3 \tilde D^c_m E^c_3.
 \label{model-4}
\end{equation}
The contributions to the soft scalar masses are given by,
\begin{eqnarray}
\delta M^2_Q &=&\left[ -\frac{\alpha_q}{24\pi}x^2 h(x) +
  \frac{\alpha_q}{16\pi^2}\left[6\alpha_q + 6\alpha_t +
\alpha_u
    -\frac{16}{3}\alpha_3 -3\alpha_2-\frac{13}{15}\alpha_1 \right]
  -\frac{\alpha _t \alpha _u}{16 \pi ^2} \right] \Lambda^2, \nonumber
\\ 
\delta M^2_{U^c} &=&\left[-\frac{2\alpha_q +\alpha_u}{24\pi}x^2
  h(x) + \frac{\alpha_q}{8\pi^2}\left(6\alpha_q + \alpha_b +
6\alpha_t
  +2\alpha_u -\frac{16}{3}\alpha_3 -3\alpha_2-\frac{13}{15}\alpha_1
  \right) \right. \nonumber \\ &&\left.+
  \frac{\alpha_u}{16\pi^2}\left( 5\alpha_u + 2\alpha_\tau
  -\frac{16}{3}\alpha_3- \frac{28}{15}\alpha_1\right) \right]
\Lambda^2,\\ \delta M^2_{D^c} &=& -\frac{\alpha_b \alpha_q}{8
  \pi^2}\Lambda^2, ~~ \delta M^2_L = -\frac{3 \alpha _u \alpha _{\tau
}}{16 \pi ^2} \Lambda^2,\nonumber \\ \delta M^2_{E^c} &=&
\left[-\frac{\alpha_u}{8\pi}x^2 h(x)
  +\frac{3\lambda_u}{16\pi^2}\left(5\lambda_u + 2\lambda_t +
  2\lambda_q -\frac{16}{3}\alpha_3- \frac{28}{15}\alpha_1 \right)
  \right] \Lambda^2,\nonumber \\ \delta M^2_{H_u} &=&
-3\alpha_t\frac{3 \alpha_q + \alpha _u}{16\pi^2}\Lambda^2,~~ \delta
M^2_{H_d} = -3\frac{\alpha_b \alpha_q + \alpha _u \alpha _{\tau
}}{16\pi^2}\Lambda^2, \nonumber \\ A_t &=& -\frac{3 \alpha _q+\alpha
  _u}{4 \pi }\Lambda, ~~ A_b = -\frac{\alpha _q}{4 \pi }\Lambda, ~~
A_\tau = -\frac{3 \alpha _u}{4 \pi } \Lambda.\nonumber
\end{eqnarray}
The lack of messenger parity in this case can lead to tree level mass
terms of the form $m'\bar{5}_H\bar{5}_m$. It is expected that $m' \sim
\mu,$ where $\mu$ is the usual dimensionful parameter in the MSSM
superpotential. This mass cannot be suppressed without suppressing the
operators in Eq.~\ref{model-4}.  The phenomenology of this model
though not identical, closely resembles the scenario studied in
\cite{Evans:2011bea}.

%%%%%%%%%%%%%
\subsubsection{Model 5}
%%%%%%%%%%%%%%
$\mathbf{10 \bar{5} \bar{5}_m}:$ This is the other possibility with
even R-parity for the messengers. One can distinguish it from Model 4
by considering a different flavor symmetry given by,
\begin{equation}
U(1)_F(5_m,\bar{5}_m,X)=(a+p,-p,-a).
\end{equation}
However note that flavor charges assigned in Table~\ref{table1} do not
distinguish between $\bar{5}_2$ and $\bar{5}_3.$ So this is not enough
to ensure that only third generation will couple strongly with the
messenger sector. This leads to considerable contributions to FCNC
that constraints the messenger scale. An additional complication is
related to the mass terms of the form $m' 5_H\bar{5}_m$ where $m'\sim
\epsilon^{p}\mu$.  Worse, the interaction can lead to rapid proton
decay and suppression of the order of $\epsilon^{5}$ is not strong
enough to comply with present experimental bounds. However one can
consider flavor symmetries which are uncorrelated to the origin of the
fermion mass hierarchy that can enable the required suppression. We
adopt the paradigm that this is possible, assuming this we can write
down the superpotential as,
\begin{equation}
 W_{int}=\lambda_q Q_3 D^c_3 H_d^m + \lambda_l Q_3L_3 D^c_m +
 \lambda_u U^c_3 D^c_3 D^c_m + \lambda_e L_3 E^c_3 H_d^m .
 \label{model-5}
\end{equation}
The contributions to the soft masses can be written as,
\begin{eqnarray} 
\delta M^2_Q &=& \left[ -\frac{x^2 h(x)}{24\pi}(\alpha_q +
  \alpha_l)+\frac{\alpha_l}{16\pi^2}\left[ 6\alpha_l + \alpha_e
    +\alpha_u + \alpha_\tau -\frac{16}{3}\alpha_3 -3\alpha_2
    -\frac{7}{15} \alpha_1 \right] \right.  \nonumber\\ 
&&\left.+   \frac{\alpha_q}{16\pi^2} \left[6\alpha_q + 6\alpha_b + 2
\sqrt{\alpha_b \alpha_\tau \alpha_e/\alpha_q} +2\alpha_l +\alpha_e
    +\alpha_u -\frac{16}{3}\alpha_3 -3\alpha_2 -\frac{7}{15} \alpha_1
    \right]-\frac{\alpha_u(\alpha_b + \alpha_t)}{16 \pi^2}\right]
\Lambda^2, \nonumber\\
\delta M^2_{U^c} &=& \left[ -\frac{x^2 h(x)}{24\pi}\alpha_u +
  \frac{\alpha_u}{16\pi^2}\left[ 3\alpha_u + 2\alpha_q + 2\alpha_l +
    2\alpha_b -8\alpha_3 -\frac45\alpha_1 \right]
  -\frac{\alpha_t(\alpha_q + \alpha_l)}{8 \pi ^2} \right] \Lambda^2,
\nonumber\\
\delta M^2_{D^c} &=& \left[ -\frac{x^2 h(x)}{24\pi}(2\alpha_q +
  \alpha_u) + \frac{\alpha_q}{16\pi^2}\left[12\alpha_q +2\alpha_l
    +2\alpha_e +2 \alpha_t + 12
\alpha_b + 4 \sqrt{\alpha_b \alpha_\tau \alpha_e/\alpha_q}
-\frac{32}{3}\alpha_3 -6\alpha_2
    -\frac{14}{15}\alpha_1 \right] \right.\nonumber\\ &&
  \left. +\frac{\alpha_u}{16\pi^2}\left[ 3\alpha_u + 2\alpha_t +
    4\alpha_q + 2\alpha_l -8 \alpha_3 -\frac45\alpha_1 \right]
  -\frac{\alpha _b \alpha _l}{8 \pi ^2} \right] \Lambda^2,\\
% % % 
\delta M^2_L &=& \left[ -\frac{x^2 h(x)}{24\pi}\alpha_e +
  \frac{\alpha_e}{16\pi^2}\left[4\alpha_e + 3\alpha_q + 2
\alpha_\tau + 6 \sqrt{\alpha_b \alpha_\tau \alpha_q/ \alpha_e}- 3\alpha_2
    -\frac95\alpha_1 \right] \right. \nonumber\\ && \left. +
  \frac{\alpha_l}{16\pi^2}\left[18\alpha_l + 6\alpha_e +3\alpha_b
    +3\alpha_q + 3 \alpha_t + 3\alpha_u -16 \alpha_3 -9\alpha_2 -
    \frac{7}{5}\alpha_1 \right] \right] \Lambda^2, \nonumber \\
\delta M^2_{E^c} &=& \left[ -\frac{x^2 h(x)}{12\pi}\alpha_e +
  \frac{\alpha_e}{8\pi^2}\left[4\alpha_e + 3\alpha_l + 3\alpha_q
    -3\alpha_2 + 2 
\alpha_\tau + 6 \sqrt{\alpha_b \alpha_\tau \alpha_q/ \alpha_e}-\frac95
\alpha_1\right]-\frac{3 \alpha_l\alpha_{\tau
  }}{8\pi^2} \right] \Lambda^2 \nonumber\\
\delta M^2_{H_u} &=& -\frac{3 \alpha_t}{16 \pi^2}(\alpha_l + \alpha_q
+ \alpha_u)\Lambda^2,~~ \delta M^2_{H_d} = -\left[ \frac{3
    \alpha_b}{16 \pi ^2} (\alpha_l+ 3\alpha_q +
  \alpha_u)+\frac{3\alpha_{\tau }}{16 \pi^2}(\alpha_e + \alpha_l)
  \right] \Lambda^2, \nonumber \\ A_t & =&-\frac{\alpha _l+\alpha
  _q+\alpha _u}{4 \pi } \Lambda,~~ A_b = -\frac{\alpha _l+3 \alpha
  _q+\alpha _u}{4 \pi }\Lambda, ~~ A_\tau = -\frac{3 \left(\alpha
  _e+\alpha _l\right)}{4 \pi }\Lambda. \nonumber
\end{eqnarray}

 Note that in  \cite{Han:1998xy} a scenario that includes Model 4 and Model 5
 was studied. Low energy flavor observables
 were utilized to constraint different cmbinations  of the interaction 
 Yukawa couplings. Considering that most flavor constraints are
 restrictive only for the first two fermion generations, the flavor symmetries 
 imposed in our analysis will certainly relax some of  these bounds. However a detailed study of these
 constraints, that warrants careful attention,
 is beyond the scope of this paper. 
 
\subsubsection{Model 6}
%%%%%%%%%%
$\mathbf{10 \bar5_m \bar5_m:}$ We expand the R-symmetry from $Z_2$ to
$Z_4$.  The parity of the MSSM multiplets are still given by
Table~\ref{table1}, whereas the messenger sector now has the
following parity,
\begin{equation}
 Rp(5_m,\bar{5}_m) = (i 5_m, -i\bar{5}_m).
\end{equation}
On top of this we impose the following flavor charges,
\begin{equation}
U(1)_F(5_m,\bar{5}_m,X)=(-a,0,a),
\end{equation}
This ensures that we have the following dominant superpotential,
\begin{equation}
W_{int}= \lambda Q_3 D^c_m H_d^m. 
\end{equation}
The new contributions are given by,
\begin{eqnarray}
\delta M^2_Q &=&\left[ - \frac{\alpha_\lambda}{12\pi} x^2 h(x) +
  \frac{\alpha_\lambda}{16\pi^2}\left(6\alpha_\lambda-
  \frac{16}{3}\alpha_3- 3 \alpha_2 -\frac{7}{15}\alpha_1 \right)
  \right] \Lambda^2 \nonumber, \\ 
\delta M^2_{U^c} &=&
-\frac{\alpha_\lambda\alpha_t}{8\pi^2}\Lambda^2,~~ \delta M^2_{D^c} =
-\frac{\alpha_\lambda\alpha_b}{8\pi^2}\Lambda^2, \\
 \delta M^2_{H_u}
&=& -\frac{3\alpha_\lambda\alpha_t}{16\pi^2} \Lambda^2,~~ \delta
M^2_{H_d} = -\frac{3\alpha_\lambda\alpha_b}{16\pi^2} \Lambda^2,
\nonumber\\ A_t&=&A_b=-\frac{\alpha_\lambda}{4\pi}\Lambda. \nonumber
\end{eqnarray}

\subsubsection{Models 7 \& 8}
%%%%%%%%%%%%%%%%%%%%%%%%
$\mathbf{10 \bar5_{jm} \bar5_{km}:}$ Sticking to the same invariant as
Model 6 an interesting scenario develops when we expand the number of
messenger generations to two. The drastic change is more than a simple
duplication of the results in Model 6. The form of the Lagrangian and
thus the soft spectrum, depends on the symmetries we impose on the
theory. In this regard we will discuss two slight variants:
 \begin{itemize}
  \item {\bf Model 7:} We consider a $Z_4$ R-parity with,
  \begin{equation}
 Rp(5_{jm},\bar{5}_{jm}) = (i 5_{jm}, -i\bar{5}_{jm}),~~j=1,2,
\end{equation}
and flavor charges,
\begin{equation} 
U(1)_F(5_{jm},\bar{5}_{jm},X)=(-a,0,a).
\end{equation}
  \item {\bf Model 8:} We consider the following  $Z_2$ R-parity,
   \begin{equation}
 Rp(5_{1m},\bar{5}_{1m},5_{2m},\bar{5}_{2m}) = (- 5_{1m},
-\bar{5}_{1m},5_{2m},\bar{5}_{2m}),
\end{equation}
 and the following flavor symmetry:
\begin{equation}
 U(1)_F(5_{1m},\bar{5}_{1m},5_{2m},\bar{5}_{2m},X)=(a-b,b,a+b,-b,-a).
\end{equation}

 \end{itemize}
With these symmetries we obtain the following form of the
superpotential for Model 7,
\begin{equation}
 W_{int} = (\lambda_{11} Q_3D^c_{1m} H_d^{1m} + \lambda_{22}
 Q_3D^c_{2m} H_d^{2m}) + \lambda_{12}Q_3D^c_{1m} H_d^{2m} +
 \lambda_{21} Q_3D^c{_2m} H_d^{1m} + \lambda_u U^c_3 D^c_{1m} D^c_{2m}
 + \lambda_e E^c_3 H_d^{1m} H_d^{2m}.
\end{equation}
The new contributions to the soft spectrum for Model 7  are given by,
\begin{eqnarray}
 \delta M^2_Q &=& \left[-\frac{x^2 h(x)}{12\pi}(\alpha_{11} +
   \alpha_{12} + \alpha_{21} + \alpha_{22}) +
   \frac{\alpha_{11}}{16\pi^2}\left[\left(3\alpha_{11} +\alpha_e +
     \alpha_u - \frac{16}{3}\alpha_3 - 3\alpha_2 - \frac{7}{15}
     \alpha_1\right)\right.\right. \nonumber \\ &&\left.+
     (\alpha_{11}\to \alpha_{12}) + (\alpha_{11}\to \alpha_{21}) +
     (\alpha_{11}\to \alpha_{22})\right]
   -\frac{\alpha_t\alpha_u}{16\pi^2} \nonumber \\ && \left.+
   \frac{1}{8\pi^2}(3\alpha_{11}\alpha_{12} + 4\alpha_{11}\alpha_{21}
   + \alpha_{11}\alpha_{22} + \alpha_{12}\alpha_{21} +
   4\alpha_{12}\alpha_{22} + 3 \alpha_{21}\alpha_{22}) \right]
 \Lambda^2, \nonumber\\
%%%
\delta M^2_{U^c} &=& \left[-\frac{x^2 h(x)}{12\pi}\alpha_u
  -\frac{\alpha_t}{8\pi^2}(\alpha_{11} + \alpha_{12} + \alpha_{21} +
  \alpha_{22}) \right. \nonumber \\ &&\left.+
  \frac{\alpha_u}{16\pi^2}\left[3\alpha_u + 2(\alpha_{11} +\alpha_{12}
    +\alpha_{21} +\alpha_{22}) -8\alpha_3 -\frac45\alpha_1 \right]
  \right] \Lambda^2,\\
%%%
\delta M^2_{D^c} &=& \left[-\frac{\alpha_b}{8\pi^2}(\alpha_{11} +
  \alpha_{12}+ \alpha_{21} + \alpha_{22}) \right] \Lambda^2,~~ \delta
M^2_L = \left[ -\frac{\alpha_e\alpha_\tau}{8 \pi ^2}\right] \Lambda^2,
\nonumber \\
\delta M^2_{E^c} &=& \left[ -\frac{x^2 h(x)}{6\pi}\alpha_e +
  \frac{\alpha_e}{8\pi^2}\left[4\alpha_e + 3 (\alpha_{11} +
    \alpha_{12}+ \alpha_{21} + \alpha_{22}) -3\alpha_2
    -\frac95\alpha_1\right] \right] \Lambda^2, \nonumber \\
\delta M^2_{H_u} &=& \left[-\frac{3\alpha_t}{16\pi^2}(\alpha_{11} +
  \alpha_{12} + \alpha_{21} + \alpha_{22} + \alpha_u) \right]
\Lambda^2,\nonumber \\ \delta M^2_{H_d} &=&\left[
  -\frac{3\alpha_t}{16\pi^2}(\alpha_{11} + \alpha_{12} + \alpha_{21} +
  \alpha_{22} + \alpha_u) -\frac{\alpha_e\alpha_\tau}{8\pi^2} \right]
\Lambda^2, \nonumber \\ A_t &=&
\left[-\frac{\alpha_{11}+\alpha_{12}+\alpha_{21}+ \alpha_{22}+\alpha
    _u}{4 \pi } \right] \Lambda,~~ A_b =
\left[-\frac{\alpha_{11}+\alpha_{12}+\alpha_{21}+\alpha_{22}}{4\pi }
  \right] \Lambda,~~ A_\tau = \left[-\frac{\alpha _e}{2 \pi } \right]
\Lambda. \nonumber
\end{eqnarray}
 
The corresponding superpotential and the soft spectrum for Model 8 can
be obtained by setting $\lambda_{11}=\lambda_{22}=0.$

\subsection{Only $10\oplus\overline{10}$}
 In this section we will collect the possible messenger-matter
 interaction terms possible assuming that the messengers are in a
 vector like representation of $10\oplus\overline{10}$.
%%%%%%%%
\subsubsection{Model 9}
%%%%%%%%%%
$\mathbf{10 10_m 5_H + 10_m 10_m 5_H:}$ Consider that the messenger
fields are odd under R-parity $(Rp(10_m)=-10_m)$ and have the following
flavor charges,
 \begin{equation}
  U(1)_x(10_m,\bar{10}_m,X)=(0,-a,a)
 \end{equation}
The invariant $10_m~\,\bar5_{2,3}~\bar5_H$ has a coupling
which is suppressed by the flavor factor and is at least as small as
the $\lambda_{\tau}$ Yukawa. We thus neglect it from the discussion
and obtain the following interaction
superpotential\cite{Albaid:2012qk},
\begin{equation}
 W_{int}=\lambda_q Q_3U^c_m H_u + \lambda_u Q^mU^c_3 H_u +\lambda_h
 Q_mU^c_m H_u.
\end{equation}
The lack of messenger parity in this scenario again manifests into one
loop mixing between messenger and matter multiplets.  The one and two
loop contributions to the soft masses including the mixing effects are
given by,
\begin{eqnarray}
\delta M^2_Q &=& \left[-\frac{\alpha_q}{24\pi}x^2 h(x)
  +\frac{\alpha_q}{16\pi^2}\left( 6\alpha_q + 3\alpha_u + 5\alpha_h +
3 \alpha_t
  -\frac{16}{3}\alpha_3 -3\alpha_2- \frac{13}{15}\alpha_1 \right)-
  \frac{\alpha_t (5\alpha_u + 3\alpha_h)}{16\pi^2}\right]
\Lambda^2,\nonumber\\
 \delta M^2_{U^c} &=& \left[-\frac{\alpha_u}{12\pi}x^2 h(x)
+\frac{\alpha_u}{8\pi^2}\left(
  6\alpha_u + 3\alpha_q + 4\alpha_h + 3 \alpha_t
-\frac{16}{3}\alpha_3 -3\alpha_2-
  \frac{13}{15}\alpha_1 \right) - \frac{\alpha_t(4\alpha_q +
    3\alpha_h)}{8\pi^2}\right] \Lambda^2,\nonumber\\ \delta M^2_{D^c}
&=& \left[-\frac{\alpha_b \alpha_q}{8 \pi^2}\right] \Lambda^2,~~
\delta M^2_{H_d} = \left[-\frac{3 \alpha_b \alpha_q}{16 \pi^2}\right]
\Lambda^2, \nonumber\\
 \delta M^2_{H_u} &=&\left[-\frac{\alpha_q +
    \alpha_u + 2\alpha_h}{8\pi}x^2 h(x) -\frac{\alpha_q+ \alpha_u +
    \alpha_h}{16\pi^2}\left(16\alpha_3 + 9\alpha_2 +
  \frac{13}{5}\alpha_1 \right) \right.\\
 &&\left. +
  \frac{3\alpha_q}{16\pi^2}\left (6\alpha_q + 10\alpha_h +\alpha_b +
5 \alpha_t   \right) + \frac{9\alpha_u}{8\pi^2}\left( \alpha_u + \alpha_q +
  \frac43 \alpha_h + \frac43 \alpha_t\right) +\frac{9
\alpha_h^2}{8 \pi ^2}\right]
\Lambda^2,\nonumber\\ A_t &=& -\frac{3 \alpha _h+4 \alpha _q+5 \alpha
  _u}{4 \pi }\Lambda, ~~A_b = -\frac{\alpha_q}{4 \pi }\Lambda.
\nonumber
\end{eqnarray}

\subsubsection{Model 10} $\mathbf{10_m\,10_m\, 5_H:}$ The other alternative is
to consider that the messengers are even under R-parity
$(Rp(10_m)=10_m)$. In this case the messengers can only couple to the
Higgs multiplets in the MSSM sector and thus we select the flavor
charges of the messenger sector to be zero.  The corresponding
superpotential is given by\cite{Craig:2012xp},
\begin{equation}
 W_{int}= \lambda Q_m U^c_m H_u,
\end{equation}
and the contributions to the soft masses are given by,
\begin{eqnarray}
\delta M^2_Q &=& -\frac{3 \alpha _t \alpha _{\lambda }}{16 \pi ^2}
\Lambda^2, ~~ \delta M^2_{U^c} = -\frac{3 \alpha _t \alpha _{\lambda
}}{8 \pi ^2}\Lambda^2, \\ \delta M^2_{H_u} &=&
\left(\frac{\alpha_\lambda}{8\pi^2}\left[9\alpha_\lambda - 8\alpha_3 -
  \frac92 \alpha_2 -\frac{13}{10}\alpha_1\right] -\frac{\alpha
  _{\lambda }}{4\pi}x^2 h(x)\right)\Lambda^2, \\ A_t &=& -\frac{3
  \alpha _{\lambda }}{4 \pi }\Lambda. \nonumber
\end{eqnarray}

\subsection{Mixed messenger models}
It is possible that the messenger sector is composed  of messenger fields
that are in different complete representations of the GUT group
SU(5). Actually this is implicitly assumed in Models 1 and 2. In that
case we can have scenarios where more than one of them simultaneously
interact with the visible sector.  There are a large number of
possibilities in this class, mainly coming from a combination of two
or  more models already discussed, possibly augmented by some new terms. A study of
all these models are beyond the scope of this study and we will
restrict ourselves to models that lead to entirely new messenger-matter
interaction terms.

\subsubsection{Model 11}
Let us consider a scenario where messengers in $1$ and $5$
representations interact simultaneously with the MSSM multiplets. The only new
interaction term that can be envisaged in this case is
$\mathbf{5_H\,\bar5_m \, 1_m}.$ In order to prevent other terms from
showing up we impose the following symmetries. Again we conjecture the
existence of a second singlet $\tilde S_m$. We impose the following
assignment of R-parity,
\begin{equation}
 Rp(5_m,\bar5_m,1_m,\tilde{1}_m)= (-5_m,-\bar5_m,-1_m,-\tilde{1}_m),
\end{equation}
and the following flavor charges,
\begin{equation}
 U(1)_F(5_m,\bar5_m,1_m,\tilde1_m,X) = (-a- b,b,-b, -a+b ,a) \mbox{
   with $a>b$}.
\end{equation}
We obtain the following unsuppressed terms in the
superpotential\cite{Craig:2012xp},
\begin{equation}
 W_{int}=H_u H_d^m S_m,
\end{equation}
and the contributions to the soft masses are given by,
\begin{eqnarray}
\delta M^2_Q &=& -\frac{\alpha_t \alpha_\lambda}{16 \pi
  ^2}\Lambda^2,~~ \delta M^2_{U^c} = -\frac{\alpha _t \alpha _{\lambda
}}{8 \pi ^2} \Lambda^2, \nonumber\\ \delta M^2_{H_u} &=& \left[
  \frac{\alpha_\lambda}{16\pi^2} \left[4 \alpha_\lambda -3\alpha_2 -
    \frac{3\alpha_1 }{5}\right] - \frac{ \alpha_\lambda }{12\pi}x^2
  h(x) \right] \Lambda^2,\\ A_t &=& -\frac{\alpha _{\lambda }}{4 \pi
}\Lambda. \nonumber
\end{eqnarray}

\subsubsection{Model 12}
The messenger sector can very well be made up of singlets and
$10\oplus \overline{10}$ multiplets. If both are allowed to couple we
can have the following new mixed invariant $\mathbf{10
  \overline{10}_m1_m}.$ In order for this to be possible we can assume
that the messenger fields have the following R-parity,
 \begin{equation}
 Rp(10_m,\overline{10}_m,1_m,\tilde{1}_m)=
 (-10_m,-\overline{10}_m,1_m,\tilde{1}_m),
\end{equation}
 and the following flavor charges,
 \begin{equation}
 U(1)_F(10_m,\overline{10}_m,1_m,\tilde{1}_m,X) = 
(a-b,b,-b,a+b,-a).
\end{equation}
The leading terms in the superpotential are given by,
\begin{equation}
W_{int}= \lambda_q Q_3\tilde Q_m S_m + \lambda_u U^c_3 \tilde U^c S_m
+ \lambda_e E^c_3 \tilde E^c_m S_m,
\end{equation}
the new contributions to the soft scalar masses are given by,
\begin{eqnarray}
\delta M^2_Q &=& \left[-\frac{\alpha_q x^2 h(x)}{12\pi}
  +\frac{\alpha_q}{16\pi^2}\left[ 8 \alpha_q + 3 \alpha_u + \alpha_e
    -\frac{16}{3}\alpha_3- 3\alpha_2 -\frac{1}{15}\alpha_1 \right]
  -\frac{\alpha_t \alpha_u}{16\pi^2}\right]\Lambda^2,\nonumber\\
\delta M^2_{U^c} &=& \left[-\frac{\alpha_u x^2 h(x)}{12\pi}
+\frac{\alpha_u}{16\pi^2}\left[6\alpha_q + 5\alpha_u + \alpha_e
-\frac{16}{3}\alpha_3 -\frac{16}{15}\alpha_1 \right] -\frac{\alpha_t
\alpha_q}{8\pi^2}\right]\Lambda^2,\nonumber\\
\delta M^2_{D^c} &=& -\frac{\alpha_b \alpha_q}{8 \pi^2}\Lambda^2,~~
\delta M^2_L = -\frac{\alpha_e \alpha_\tau}{16\pi^2} \Lambda^2,\nonumber \\
\delta M^2_{E^c} &=& \left[-\frac{\alpha_e x^2 h(x)}{12\pi}
+\frac{3\alpha_e}{16\pi^2}\left[\alpha_e + 2\alpha_q + \alpha_u  
-\frac{4}{5}\alpha_1\right]\right]\Lambda^2,\\
\delta M^2_{H_u} &=& -\frac{3 \alpha_t(\alpha_q + \alpha_u) }{16\pi^2}
\Lambda^2,~~ \delta M^2_{H_d} = -\frac{3 \alpha_b \alpha_q +
  \alpha_e\alpha_\tau}{16\pi^2}\Lambda^2, \nonumber\\ A_t &=&
-\frac{\alpha_q + \alpha _u}{4 \pi } \Lambda,~~ A_b = -\frac{\alpha
  _q}{4 \pi }\Lambda,~~ A_\tau = -\frac{\alpha _e}{4 \pi
}\Lambda.\nonumber
\end{eqnarray}

\subsubsection{Models 13 \& 14}
Consider scenarios where the messenger sector is made up of vector
pairs of $5\oplus\bar5$ and $10\oplus\overline{10}$.  There are two
distinct new interaction terms that can arise other than combinations
of the models already studied earlier. We are going to consider these
two model one by one,
 
$\mathbf{10 10_m 5_m:}$ One can motivate this by considering the
following symmetry arrangements. Consider the following R-parity,
 \begin{equation}
  Rp(5_m,\bar5_m,10_m,\overline{10}_m)=(-5_m,-\bar5_m,10,\overline{10}_m),
 \end{equation}
and the following flavor charges,
\begin{equation}
 U(1)_F(5_m,\bar5_m,10_m,\overline{10}_m,X)=(b, a-b,-b,a+b,-a),
\end{equation}
with $a>>1$ as always. Thus the part of the superpotential that
remains unsuppressed by the flavor factors is given by,
\begin{equation}
W_{int}= \lambda_{q1} Q_3 U^c_m H_u^m +  \lambda_{u1} Q_m U^c_3 H_u^m +
\lambda_{q2} Q_3 Q_m \tilde D^c_m + \lambda_{u2} U_3^c E^c_m \tilde D^c_m + 
\lambda_e U_m^c E^c \tilde D^c_m , 
\end{equation}
Note that  proton decay can occur  
at one loop through the interactions in the superpotential.  The  suppression 
by the Froggatt-Nielsen factor notwithstanding, it requires a severe fine-tuning of the 
superpotential parameters to be consistent with proton decay constraints. Assuming additional
discrete symmetry can ameliorate this problem, 
the new contributions to the soft masses are given by,
\begin{eqnarray}
\delta M^2_Q &=&\left[ -\frac{\alpha_{q1}  +\alpha_{q2}}{12\pi}x^2 h(x)
+\frac{\alpha_{q1}}{16\pi^2}\left(6\alpha_{q1}+ \alpha_e + 3 \alpha_{u1}
-\frac{16}{3} \alpha_3 -3\alpha_2 -\frac{13}{15}\alpha_3\right)\right. \nonumber \\
&&\left. + \frac{\alpha_{q2}}{16\pi^2}\left(4\alpha_{q2} + 2\alpha_{q1} +
\alpha_{u1} +\alpha_{u2} +\alpha_e -8\alpha_3 -3\alpha_2 -\frac15 \alpha_1
\right) -\frac{\alpha_t}{16\pi^2}(2\alpha_{u1}+ \alpha_{u2})\right]\Lambda^2, \nonumber \\
\delta M^2_{U^c} &=&\left[ -\frac{2\alpha_{u1}+\alpha_{u2}}{12\pi} x^2 h(x)
+\frac{\alpha_{u1}}{8\pi^2}\left(6\alpha_{u1} + 3\alpha_{q1} + \alpha_{q2}
-\frac{16}{3}\alpha_3 -3\alpha_2 -\frac{13}{15} \alpha_1 \right)\right. \nonumber \\
&&\left. + \frac{\alpha_{u2}}{16\pi^2}\left(5\alpha_{u2} + 4\alpha_{u1} +
2\alpha_{q2} + \alpha_e -\frac{16}{3}\alpha_3 -\frac{28}{15}\alpha_1\right)
-\frac{\alpha_t}{8\pi^2}(\alpha_{q1} + \alpha_{q2})\right]\Lambda^2, \nonumber \\
\delta M^2_{D^c} &=& -\frac{\alpha_b(\alpha_{q1}+
\alpha_{q2})}{8\pi^2}\Lambda^2,~~
\delta M^2_L = -\frac{3 \alpha_e \alpha_{\tau }}{16 \pi ^2}\Lambda^2, \nonumber \\
\delta M^2_{E^c} &=& \left[-\frac{\alpha_e}{4\pi} x^2 h(x)
+\frac{3\alpha_e}{16\pi^2}\left(5\alpha_e + 2(\alpha_{q1} + \alpha_{q2})
+\alpha_{u2}-\frac{16}{3}\alpha_3- \frac{28}{15}\alpha_1
\right)\right]\Lambda^2, \\
\delta M^2_{H_u} &=& -3 \alpha_t\frac{\alpha_{q1} + \alpha_{q2} + 2
\alpha_{u1} + \alpha_{u2}}{16 \pi ^2}\Lambda^2,~~
\delta M^2_{H_d} = -3\frac{\alpha_b(\alpha_{q1}+ \alpha_{q2}) + \alpha_e
\alpha_{\tau}}{16 \pi ^2}\Lambda^2, \nonumber\\
 A_t &=& -\frac{\alpha _{q1}+\alpha _{q2}+2
\alpha_{u1}+\alpha _{u2}}{4 \pi }\Lambda,~~
 A_b = -\frac{\alpha _{q1}+\alpha _{q2}}{4 \pi } \Lambda,~~
 A_{\tau } = -\frac{3 \alpha _e}{4 \pi }\Lambda. \nonumber
\end{eqnarray}

$\mathbf{\overline{10}_m 5_m 5_H:}$ In this case consider the
following R-parity,
  \begin{equation}
  Rp(5_m,\bar5_m,10,\overline{10}_m)=(5_m,\bar5_m,10_m,\overline{10}_m),
 \end{equation}
and the following flavor charges,
\begin{equation}
 U(1)_F(5_m,\bar5_m,10_m,\overline{10}_m,X)=(b,a-b,a+b,
-b,-a).
\end{equation}
 The corresponding superpotential is given by,
 \begin{equation}
 W_{int}= \lambda_q \tilde Q_m \tilde D^c_m H_u +\lambda_e \tilde
 E^c_m H_u^m H_u,
\end{equation}
which results in the following new contributions to the soft masses,
\begin{eqnarray}
 \delta M^2_Q &=& -\frac{\alpha_t(\alpha_e + 3\alpha_q)}{16 \pi^2}
 \Lambda^2,~~ \delta M^2_{U^c} = -\frac{\alpha_t(\alpha_e +
   3\alpha_q)}{8 \pi^2} \Lambda^2,\nonumber\\ \delta M^2_{H_u} &=&
 \left[- \frac{3\alpha_q+ \alpha_e}{12\pi} x^2 h(x) +
   \frac{\alpha_e}{16\pi^2}\left( 4\alpha_e -3\alpha_2 -\frac95
   \alpha_1 \right)\right. \nonumber \\ &&\left.+
   \frac{\alpha_q}{8\pi^2} \left( 9 \alpha_q + 3\alpha_e -8 \alpha_3
   -\frac92\alpha_2 - \frac{7}{10}\alpha_1 \right)
   \right]\Lambda^2,\\ A_t &=& -\frac{\alpha_e+3 \alpha_q}{4 \pi }
 \Lambda \nonumber.
\end{eqnarray}

\section{Comparison and Results}
%%%%%%%%%%%%%%%%%%%%%%%%%%%%%%%%%%%%%%%%%%%%%%%%%%%%%%%
\begin{table}
 \begin{center}
\begin{tabular}{|c|c|c|c|c|c|}
 \hline Model $\vphantom{\frac{\frac{\frac12}{1}}{1}}$ & Main
 constraint & Analytic limit & $m_{\tilde t_1}|_{\min}$ & NLSP & Ref.
 \\ $\vphantom{\frac{1}{\frac{\frac12}{1}}}$& on $A_t$ & on
$\frac{|A_t|}{M_{\tilde g}}|_{\rm max}$ & TeV & &
 \\ \hline\hline 1
 $\vphantom{\frac{\frac{\frac12}{1}}{\frac{\frac12}{1}}}$& $M_{\tilde
   E}^2$ & $ \frac35 \frac{\alpha_1^2}{\alpha_\tau \alpha_3}$ & $2.2$
 & $\chi_1^0,\,\, \tilde \tau_1$ & \cite{Kang:2012ra}\\ \hline 2
 $\vphantom{\frac{\frac{\frac12}{1}}{\frac{\frac12}{1}}}$& $M_{\tilde
   E}^2$ & $ \frac35 \frac{\alpha_1^2}{\alpha_\tau \alpha_3}$ & $1.9$
 & $\chi_1^0,\,\, \tilde \tau_1$ & New \\ 
\hline
 3  $\vphantom{\frac{\frac{\frac12}{1}}{1}}$& $M_{\tilde L}^2$ & $
 \frac{1}{\alpha_\tau \alpha_3} \left[\frac{\alpha_2^2}{2} +
   \frac{\alpha_1^2}{10}\right]$ & $0.6$ & $\chi_1^0$, $\tilde
 \tau_1$, $\tilde\mu_1$ & \cite{Albaid:2012qk,Han:1998xy,Martin:2012dg}$^*$ \\
\hline 4
 $\vphantom{\frac{\frac{\frac12}{1}}{\frac{\frac12}{1}}}$& $M_{\tilde
   D^c}^2$, $M_{\tilde L}^2$ & $ \frac{\alpha_1^2}{5\alpha_3}\left(
 \frac{1}{\alpha_b} + \frac{1}{2\alpha_\tau}\right) $ & $ > 3$ &
 $\chi_1^0$, $\tilde \tau_1$, $\tilde\nu_e$& \cite{Han:1998xy,Evans:2011bea}$^*$
 \\ && $+ 4\frac{\alpha_3}{\alpha_b} + \frac{\alpha_2^2}{2\alpha_\tau
   \alpha_3} $ &&&\\ 
\hline 
5  $\vphantom{\frac{\frac{\frac12}{1}}{\frac{\frac12}{1}}}$& unconstrained &
$----$ & $ 0.1$ & $\chi_1^0$, $\tilde\mu_1$, $\tilde\nu_e$&
\cite{Han:1998xy}$^*$ \\
 \hline 
6  $\vphantom{\frac{\frac{\frac12}{1}}{\frac{\frac12}{1}}}$& $M_{\tilde
   U_3^c}^2$ & $ \frac{4}{3 } \left[\frac{\alpha_3}{ \alpha_t} +
   \frac{\alpha_1^2}{5\alpha_3 \alpha_t} \right]$ & $ 0.7$ & $\tilde
 \tau_1$, $\tilde\mu_1$& New \\ \hline 7
 $\vphantom{\frac{\frac{\frac12}{1}}{\frac{\frac12}{1}}}$& unconstrained &
 $----$ & $ 0.5$ & $\tilde \chi^0_1$, $\tilde\mu_1$& New \\ \hline 8
 $\vphantom{\frac{\frac{\frac12}{1}}{\frac{\frac12}{1}}}$& unconstrained &
 $----$ & $ 0.5$ & $\tilde \chi^0_1$, $\tilde\mu_1$ & New \\ \hline 9
 $\vphantom{\frac{\frac{\frac12}{1}}{\frac{\frac12}{1}}}$& unconstrained &
 $----$ & $ 0.5$ & $\tilde \tau_1$, $\tilde\chi_1^0$ &
 \cite{Albaid:2012qk,Martin:2012dg}$^*$ \\ \hline 10
 $\vphantom{\frac{\frac{\frac12}{1}}{\frac{\frac12}{1}}}$& $M_{\tilde
   U^c}^2$ & $ \frac{4}{3 } \left[\frac{\alpha_3}{ \alpha_t} +
   \frac{\alpha_1^2}{5 \alpha_3 \alpha_t} \right]$ & $ 0.6$ & $\tilde
 \tau_1$& \cite{Craig:2012xp} \\ \hline 11
 $\vphantom{\frac{\frac{\frac12}{1}}{\frac{\frac12}{1}}}$& $M_{\tilde
   U^c}^2$ & $ \frac{4}{3 } \left[\frac{\alpha_3}{ \alpha_t} +
   \frac{\alpha_1^2}{5 \alpha_3 \alpha_t} \right]$ & $ 1.9$ & $\tilde
 \tau_1$ & \cite{Craig:2012xp}\\ \hline 12
 $\vphantom{\frac{\frac{\frac12}{1}}{\frac{\frac12}{1}}}$& REWSB &
 $----$ & $ 2.9$ & $\tilde \chi^0_1$ & New \\ \hline 13
 $\vphantom{\frac{\frac{\frac12}{1}}{\frac{\frac12}{1}}}$& unconstrained
 & $ ----$ & $0.8$ & $\tilde
 \tau_1$, $\chi_1^0$, $\tilde\mu_1$, $\tilde\nu_1 $ & New \\ \hline 14
 $\vphantom{\frac{\frac{\frac12}{1}}{\frac{\frac12}{1}}}$& $M_{\tilde
   U^c}^2$ & $ \frac{4}{3 } \left[\frac{\alpha_3}{ \alpha_t} +
   \frac{\alpha_1^2}{5 \alpha_3 \alpha_t} \right]$ & $ 1.2$ & $\tilde
 \tau_1$ & New \\ \hline \hline
\end{tabular}
  \end{center}
\caption{ \em \small We summarizer the main theoretical limit on the
  generated top trilinear coupling where they are analytically
  possible. We also list the smallest value of the lightest stop
  ($m_{\tilde t_1}|_{\min}$) for which the Higgs mass could be boosted
  to the range $123-127$ GeV as obtained from our numerical
  scanning. Finally the NLSP for each model is listed. We also list
  the references for models which have been studied previously in the
  literature.  *  indicates that these models are slight
  variants of the references given or we have updated the older
  analysis to include all two loop corrections. Also note that the eventually the 
  'unconstrained' are limited by renormalization group running effects.}
\label{table2}
\end{table}
%%%%%%%%%%%%%%%%%%%%%%%%%%%%%%%%%%%%%%%%%%%%%%%%%%%%%%%%%%%%%%%

In this section we will compare the models introduced earlier.
The ability of these models to raise the Higgs mass,
  without requiring a large stop mass, is through a sizable top
  trilinear coupling $A_t$. However it is clear from the expressions
for the soft spectrum given in Eqs.~\ref{one_loop} and \ref{two_loop}
that the terms that lead to a large $A_t$ are also responsible for
tachyonic contribution to the soft masses. The size of the trilinear
coupling is thus constrained by the condition that all sfermion masses
should remain positive and that proper radiative electroweak symmetry
breaking (REWSB) should take place. In Table~\ref{table2} we
list the models along with the sfermions that are most vulnerable to the
tachyonic contributions. Assuming that the one loop tachyonic contributions are
negligible owing to the extra suppression from the messenger breaking
scale, it is possible to put upper limit on the size of the trilinear
couplings. We exhibit these upper limits for models wherever they are
analytically possible by relating them to the gluino mass, which can
be considered as the representative of the soft masses. Some observations
become apparent from the table,
\begin{enumerate}
 \item The right handed sleptons get the smallest contribution from
   gauge interactions. Thus in models where they receive negative
   contributions from the messenger-matter interactions they become
   susceptible to turn tachyonic. This constrains the size of the
   trilinear couplings in these models and thus effects their  efficiency
   in raising the Higgs masses.    \item Radiative
   electroweak symmetry breaking either requires a large stop mass that
   can turn one of the neutral Higgs eigenvalue negative radiatively
   or a contribution from the messenger-matter interactions that
   drives it to a negative value. An analytic study of each of these
   models in terms of electroweak symmetry breaking is difficult and
   we will rely on a numerical simulation of these models for this
   purpose.
\end{enumerate}

The discussion above is confined to the spectrum at the supersymmetry
breaking scale which already gives us an insight to models in terms of
their ability to generate large trilinear couplings and thus ease the
fine-tuning in the models.  We point out
that a generation of a large trilinear is not enough to
alleviate the problem with the Higgs mass. In principle the subsequent
renormalization group evolution to low energies can wash out the
trilinear coupling generated at the supersymmetry breaking scale due
to the partial cancellation between the Yukawa and gaugino
contributions to the $\beta$ function \cite{Martin:1997ns}. In fact a
large $A_t$ at high scale is useful if it has a sign opposite to that
of the gaugino masses. In the class of models studied here, this is
naturally achieved as is evident from Eq.~\ref{usual} and
Eq.~\ref{one_loop}. The sign difference between the trilinear
couplings and the gaugino masses can be traced back to the overall
negative factor for fermionic loops in the leading contribution to the
trilinear coupling. This generic feature of this class of models is
aided by our choice of a low scale of supersymmetry breaking which
reduces any potential harmful effect of the renormalization group
running by shortening the range.

Next we perform an extensive numerical study of each of the above
models to make a numerical comparison between them. The numerical
procedure that is followed for each of these models is described
below:
\begin{enumerate}
 \item We consider $\Lambda=F/M$ and $x=\Lambda/M$ as the two
   independent parameters that define all the scales in the
   theory. In order to scan over the parameter space of the models we
   vary $\Lambda$ between $4\times 10^4 - 9\times 10^5$ GeV. We
   consider two different values of $x\sim .01,.1$ to include
   scenarios where the one loop contributions become insignificant and
   comparable to the corresponding two loop contributions respectively.
 \item  
We use the known Standard Model fermion masses at the weak scale and
$\tan\beta$ as the boundary conditions and use one loop renormalization
group equations to determine the known coupling constant at the
messenger scale defined by $M_{mess}=\Lambda/x$.
\item 
For given values of $\Lambda$,  $x$, MSSM couplings and a given set of
the new messenger-matter Yukawas we generate the soft spectrum at the
messenger scale.
 \item We then use one loop renormalization group equations
   \cite{Martin:1997ns} to determine the weak scale spectrum. Within
   the framework of pure gauge mediation the $\mu$ term in the MSSM
   superpotential cannot be generated.  Note that messenger-matter
   interactions can in principle  lead to generation of a $\mu$
   term at the messenger scale, see
   \cite{Craig:2012xp,Delgado:2007rz}. However all  the models
   studied in this paper except Model 1 do not give rise to a $\mu$
   term\footnote{ However generation of the  $\mu$ term through gauge
     mediation can be incorporated in these models through the usual
     extension to a $Z_3$ NMSSM like scenario
     \cite{Craig:2012xp}.}. In our numerical study we will simply
   impose the matching condition at the weak scale for the potential
   minima and set the weak scale $\mu$ parameter from the following
   relation,
 \begin{equation}
  m_Z^2 \simeq-2(m_{H_u}^2 + |\mu|^2) +
  \frac{2}{\tan^2\beta}(m_{H_d}^2 - m_{H_u}^2 ).
 \end{equation}
 \item Finally we use SuSpect \cite{Djouadi:2002ze} and micrOMEGAS
   \cite{Belanger:2010gh,Belanger:2004yn} to determine correct electroweak
symmetry breaking, diagonalize the sparticle mass matrices and determine
   weak scale observables.
 \item We scan over all the new messenger-matter couplings $\lambda_i$
   between $.05-1.5$.  For every choice of the Yukawa couplings we
   repeat the above procedure to scan over the parameter space of the
   models.
\end{enumerate}

\begin{figure*}[h]
\centering \subfigure[Only Singlets]
           {\includegraphics[width=.34\textwidth,keepaspectratio,angle=270]
             {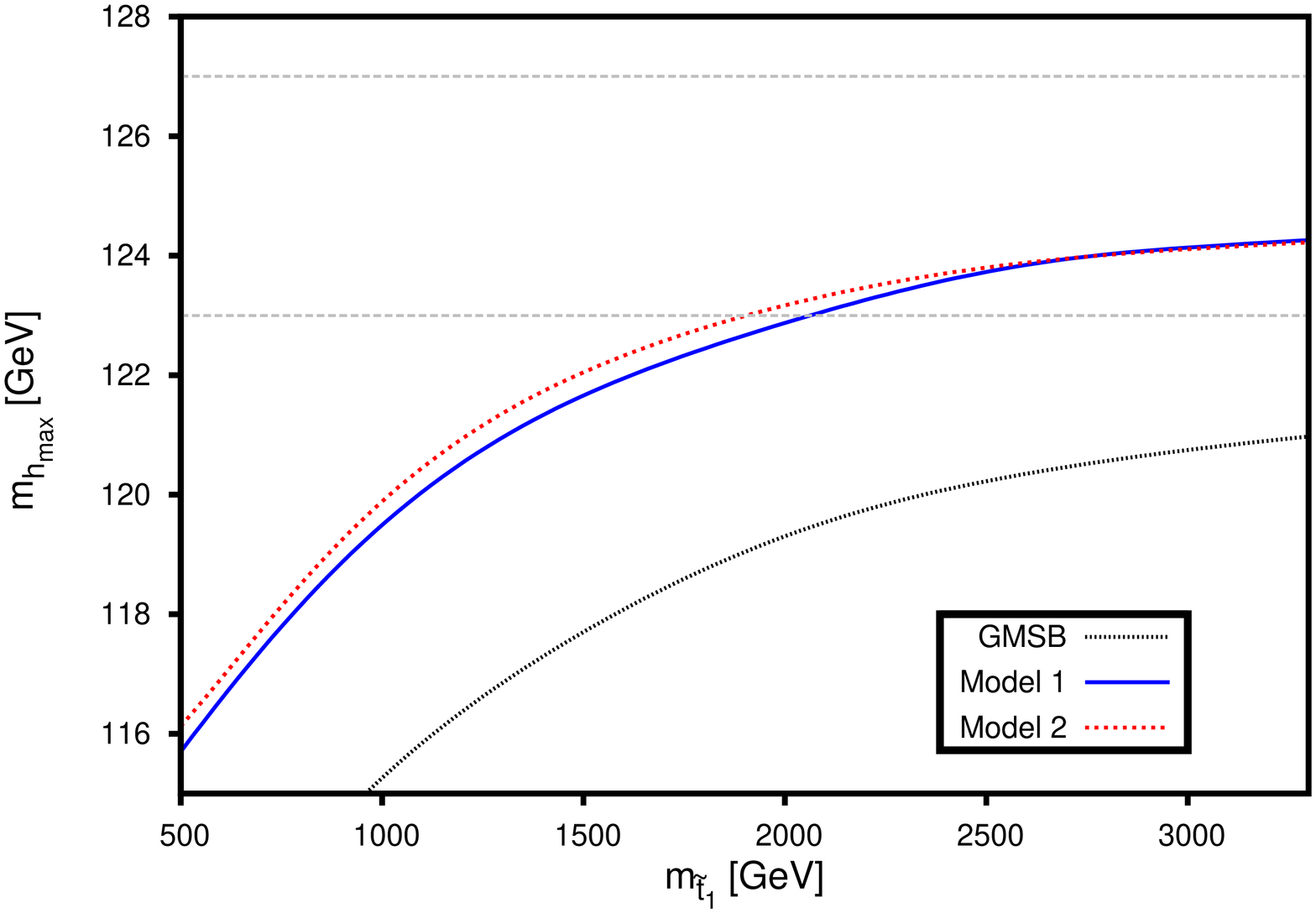}} \subfigure[Only
             $5\oplus\bar{5}$]{\includegraphics[width=.34\textwidth,keepaspectratio,
               angle=270 ] {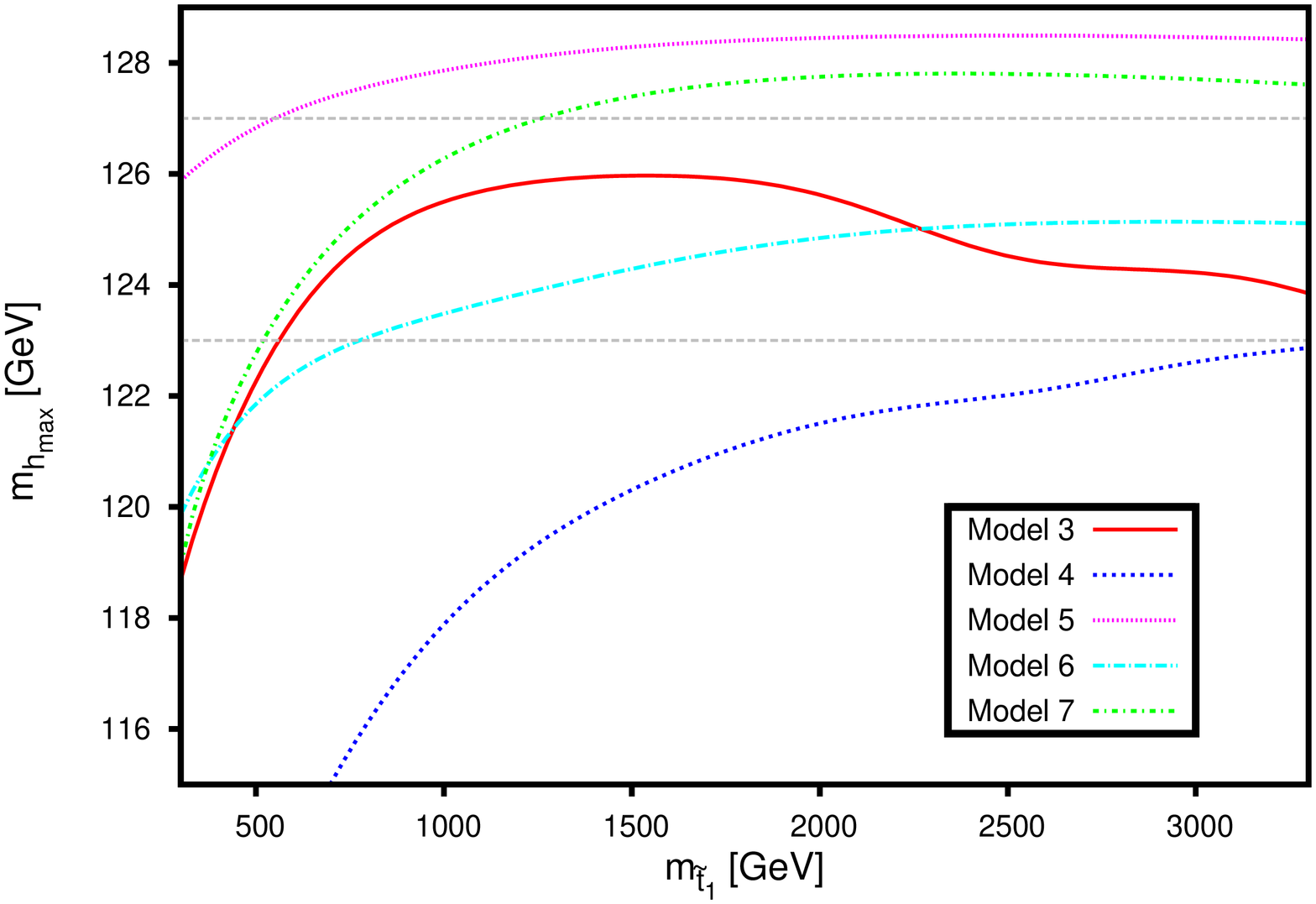}} \\ \subfigure[Only
             $10\oplus\overline{10}$]
           {\includegraphics[width=.34\textwidth,keepaspectratio,angle=270]
             {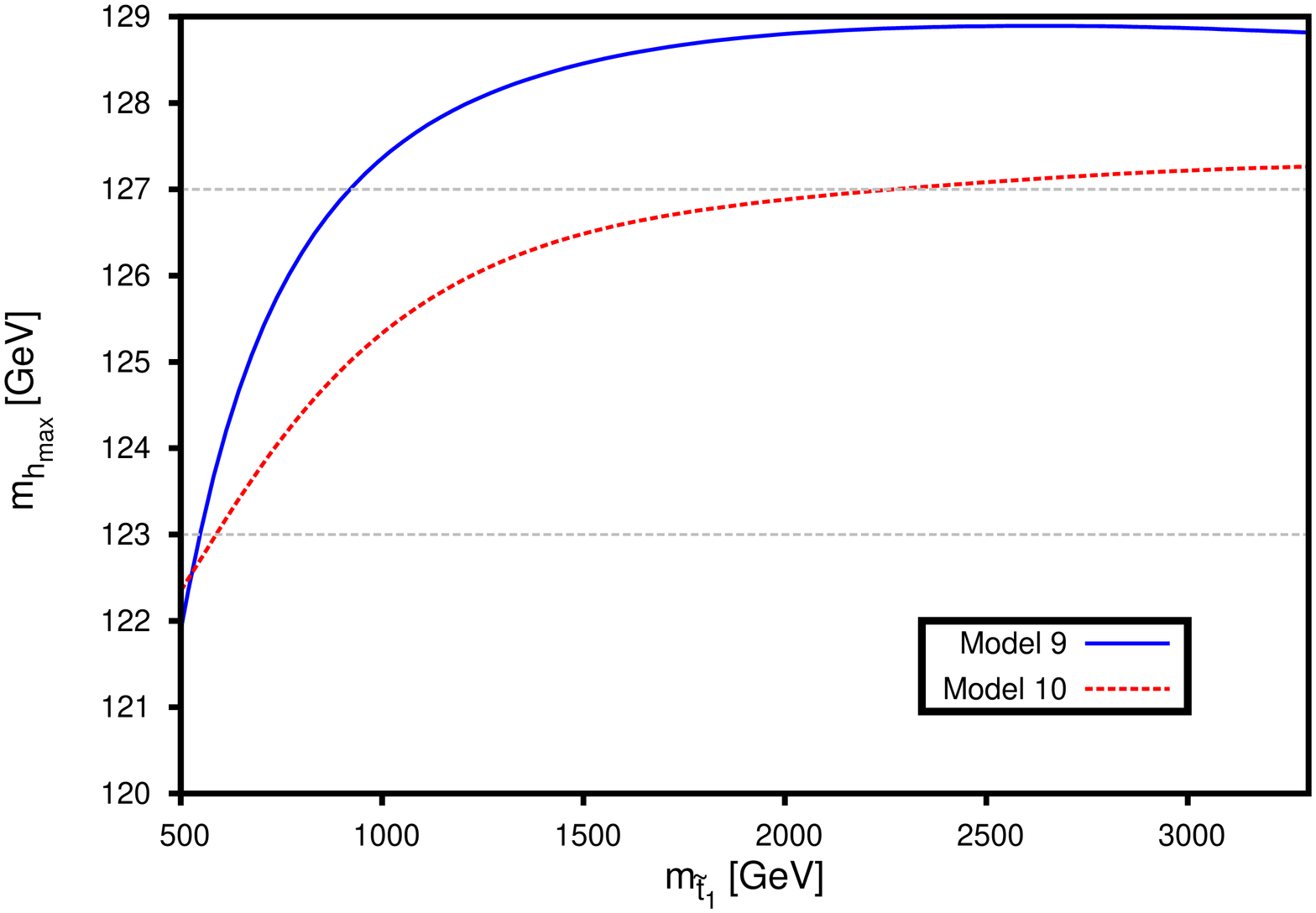}} \subfigure[Mixed messenger
             models]{\includegraphics[width=.34\textwidth,keepaspectratio,angle=270]
             {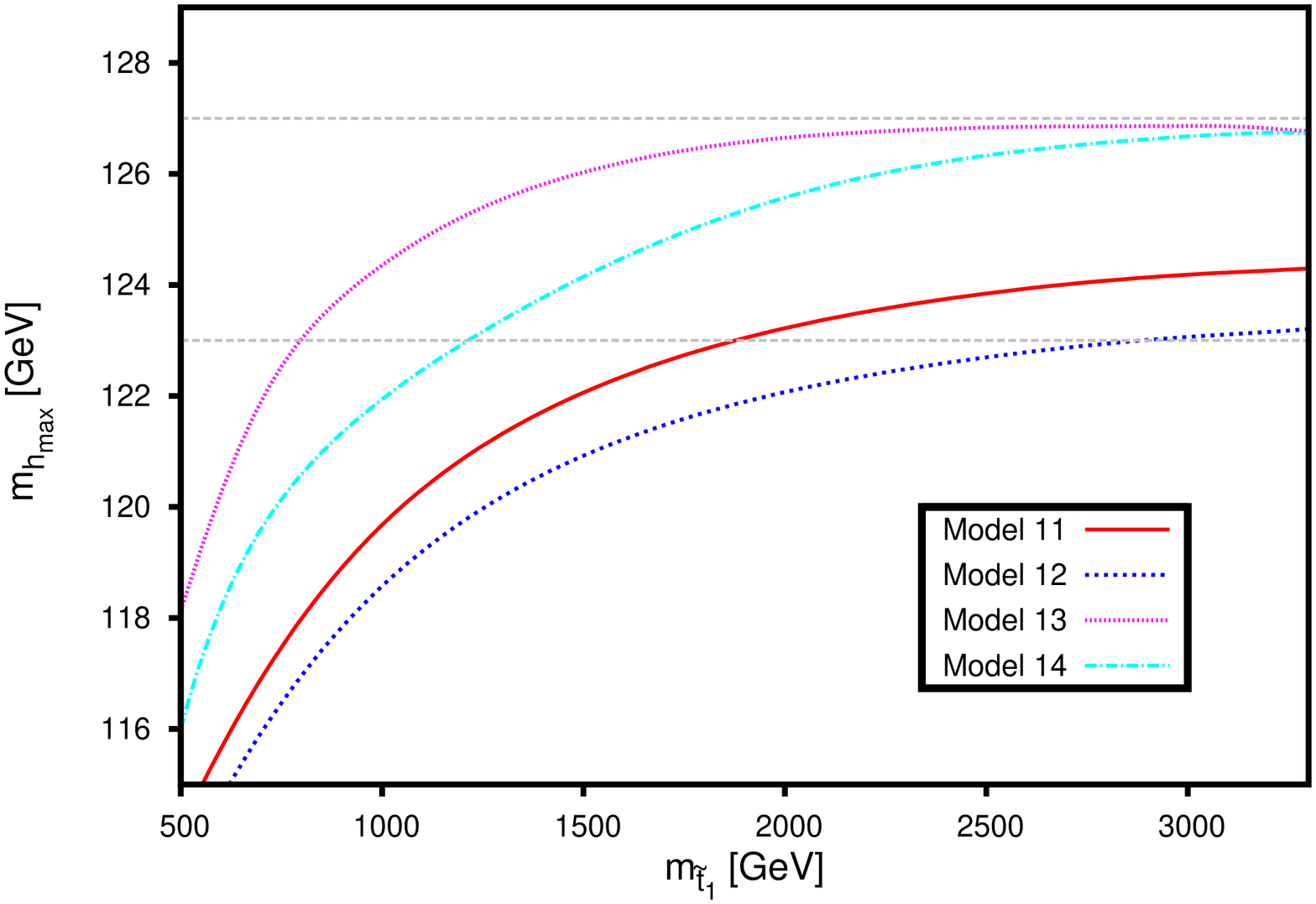}}
             
                     \caption[]{\em \small The maximum Higgs mass for
                       the different models is plotted against the
                       lightest stop mass. In all the models the new
                       Yukawa couplings are varied between $0.1-1.5$,
                       $\tan\beta$ between $5-50$, $\Lambda$ between
                       $2\times 10^4 - 9 \times 10^5$ and two
                       different choices of $x\def \Lambda/ M_{mess} =
                       .01~\&~.1$ were considered. The region below
                       each curve is accessible to the corresponding
                       model. 'GMSB' in (a) represents the mGMSB model
                       with a single messenger pair in $5 \oplus
                       \overline{5}$ of SU(5).}
\label{fig1}
\end{figure*}

In Figure~\ref{fig1} we correlate the maximum attainable Higgs mass to
the lightest stop mass for the various Models discussed in
Section~\ref{models}. We take care to sum over all model parameters
and extract the outer envelop of the valid model points to obtain the
displayed curves.  The horizontal band corresponds to the presently
preferred range of Higgs mass measured at the LHC.
This allows a
direct comparison between the models introduced in the previous
section. To reduce clutter we have not displayed the plot for Model 8
which is a subset of the Model 7 in the figures. As
anticipated from Table~\ref{table2}, models where the size of the trilinear
coupling is
constrained at the messenger scale show marginal improvement over the pure
GMSB models.  Considerations of radiative electroweak symmetry
breaking also enter the game.  The table  indicates the
smallest stop mass for which we obtain a Higgs in the desired mass range 
for every model. A
discussion about the correlation between the constraints on the trilinear
couplings and the lightest stop mass is complicated by the effects of the
renormalization group running. However a couple of comments regarding some of the
apparent contradictions in  
Table~\ref{table2} are in order: (i) Note that though the limit on the 
trilinear coupling is more relaxed in Model 4 as compared to Model 3, the 
former receives large positive two loop contributions to $M_{\tilde Q}^2$
and $M_{\tilde{U}^c}^2$ thus making the stop spectrum heavier. This should
be contrasted with the negative contribution to $M_{\tilde{U}^c}^2$ in Model 3.
However the numerically dominant effect comes from the purely negative contribution
to $M_{H_d}^2$ in Model 4 that inhibits successful REWSB in large region of the 
parameter space. (ii) While the trilinear couplings in Models 10 and
11 have the same limits they yield very different low energy spectrum. This is again
related to the difference in the contributions to $M_{H_u}^2$ resulting a more 
stringent constraint from REWSB for Model 11 as compared to Model 10. Note that
Model 14 effectively combines Models 10 and 11, however a larger Dynkin index makes
the spectrum relatively heavier.

 %%%%%%%%%%%%%%%%%%%%%%%%%%%%%%%%%%%%%%%%%%%%%%%%%%%%%%
 \begin{table}
 \begin{center}
 \begin{tabular}{|c|c|c|c|c|}
 \hline\hline Model & 5&7&9&10\\
 \hline \hline Messenger &
 $\lambda_q=1.05,$& $\lambda_{11}=\lambda_{22}=0$,& $\lambda_q=.25,$&
 $\lambda_q=1.15$\\ -matter & $ \lambda_u=1.25,$ &
 $\lambda_{12}=\lambda_{21}=1.3,$ & $\lambda_u=.75,$& \\
 Yukawa & $ \lambda_l=1.05, \lambda_e=.45$ & $\lambda_u=1.05,
 \lambda_e=.05$&$\lambda_h=1.05$&\\
 \hline
 $\Lambda$  $\vphantom{\frac{\frac12}{\frac12}}$& $1.4\times10^5$ &
 $8\times10^4$& $4\times10^4$& $7\times10^4$\\
 \hline
 \hline $M_h$  $\vphantom{\frac{\frac12}{\frac12}}$& 127 & 126 & 125
 &125\\
 \hline
 $M_H$ &2792 &1147 &257 & 829\\
 \hline 
$M_A$  &2792&1147& 257 &829\\
 \hline 
$M_{H^+}$  $\vphantom{\frac{\frac12}{\frac12}}$ & 2793 & 1150& 271&
 733\\
 \hline \hline
 NLSP $\vphantom{\frac{\frac12}{\frac12}}$&  $\chi^0_1$ & $\tilde\mu_1$
&$\tilde\tau_1$ &$\tilde\tau_1$\\
 \hline
 $m_{3/2}$ & 29.8 { eV} & 9.7 { eV} & 2.4 { eV}& 7.5 {  eV}\\
 \hline\hline
 $M_{\chi^0_1}$, $M_{\chi^0_2}$  $\vphantom{\frac{\frac12}{\frac12}}$& 190,
376 & 216, 429 &  123,171 & 283,550\\
 \hline 
$M_{\chi^0_3}$, $M_{\chi^0_4}$  $\vphantom{\frac{\frac12}{\frac12}}$&2923,
2924 &1954, 1956 &  185, 349 &753, 770 \\
 \hline 
$M_{\chi^\pm_1}$,  $M_{\chi^\pm_2}$ $\vphantom{\frac{\frac12}{\frac12}}$ & 376
,  2925 & 429 , 1956 & 149, 349 & 550, 770\\
 \hline
 $M_{\tilde g}$ & 1298 & 1491 & 1118 & 1947 \\
 \hline\hline
 $M_{\tilde t_1}$, $M_{\tilde t_2}$  $\vphantom{\frac{\frac12}{\frac12}}$&
766, 2374& 879, 1741 &  498, 850 & 985, 1557\\
 \hline
$M_{\tilde b_1}$, $M_{\tilde b_1}$ $\vphantom{\frac{\frac12}{\frac12}}$&
2363, 3142 &  1503,1726 & 732, 1041 & 985, 1557\\
 \hline 
$M_{\tilde u_R}$, $M_{\tilde u_L}$& 1635, 1719 & 1557, 1600 &
 1061,1087& 1849,1895 \\
 \hline $M_{\tilde d_R}$, $M_{\tilde d_L}$ & 1646, 1721 & 1541, 1602&
1054, 1090 &  1832,1897\\
 \hline 
$M_{\tilde\tau_1}$, $M_{\tilde\tau_2}$  $\vphantom{\frac{\frac12}{\frac12}}$ &
1342, 3347& 128, 479&  78, 294 & 56, 522\\
 \hline\hline
$\vphantom{\frac{\frac12}{\frac12}}$ Fine-tuning ($\frac{\Delta}{
\Delta_{mGMSB}}$)& 0.11  &0.10 &  0.03 & 0.10\\
 \hline\hline
\end{tabular}
\end{center}
\caption{ \em \small Representative spectrum for the chosen
  models. All the masses except the gravitino mass are given in the
  GeV unit. Here $\tan\beta=10$ and $x=0.1$ The spectrum is chosen
  conservatively to be consistent with latest experimental bounds from
  direct collider studies and low energy observables.}
\label{table3}
\end{table}

%%%%%%%%%%%%%%%%%%%%%%%%%%%%%%%%%%%%%%%%%%%%%%%%%%%%%%%%%%%%%%%%%%%%%%%%%%%%%%

\begin{figure*}
 \begin{center}
 \includegraphics[width=0.8\textwidth,angle=0,keepaspectratio]{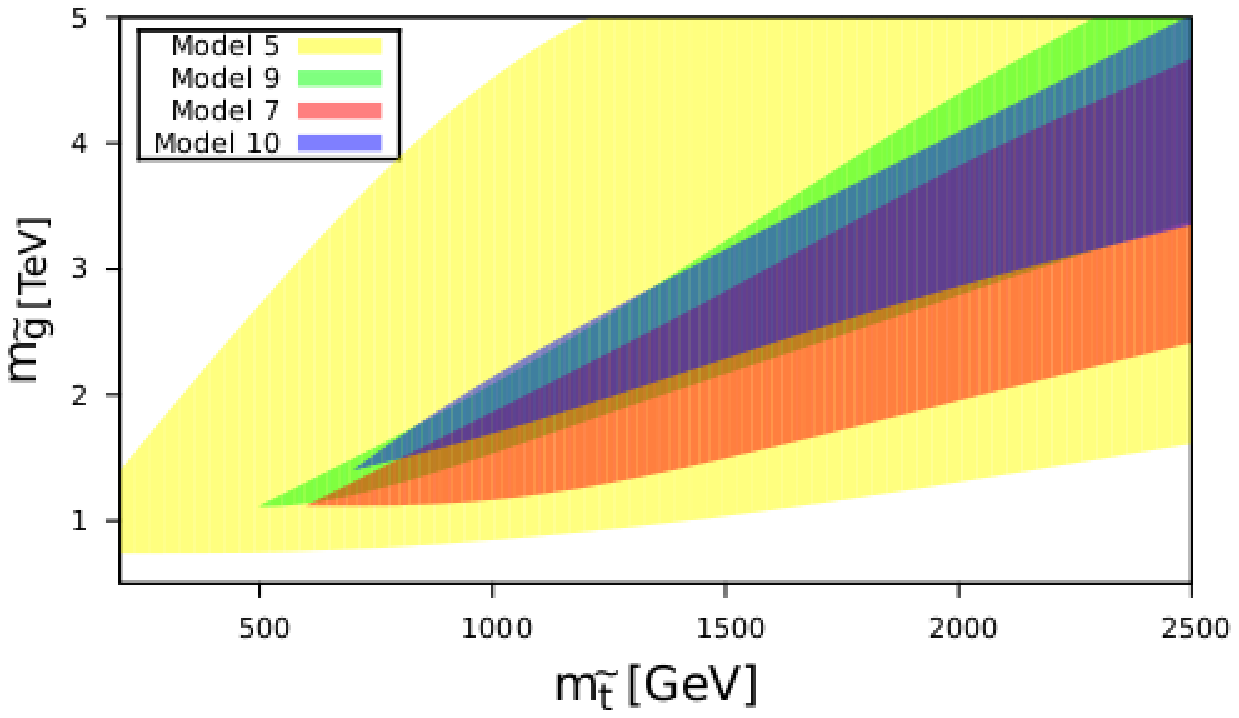}
 \end{center}
 \caption{\em \small Allowed region plot of the preferred models which
   correspond to $m_h$ between $123 -127$ GeV in the plane of the
   lightest stop mass and the gluino mass.}
\label{fig2}
\end{figure*}

Clearly
from this Models 5, 7 ,9, and 10 can be identified as most promising
in terms of boosting the Higgs mass to experimentally favored range
without admitting too much fine-tuning.  In Figure~\ref{fig2} we show
the allowed region for the Models 5, 7 ,9 and 10 where the Higgs mass
is between $123-127$ GeV, in the parameter space of the lightest stop
and the gluino masses.
These plots can be
directly translated to a measure of naturalness of the models by
relating them to the \textit{Barbieri-Giudice} \cite{Barbieri:1987fn}
fine-tuning parameter using the approximate relation
\cite{Wright:1998mk}, $\Delta \simeq {\mathcal{O}}(1) 10 t/33 (\lambda_t M_{S}/650
GeV)^2,$ where $t= \log[M_{mess}/M_Z]$, $\lambda_t$ is the top Yukawa
coupling and $M_S = \sqrt{m_{\tilde t_1}m_{\tilde t_2}}$.  A recent
study in pure mGMSB models \cite{Ajaib:2012vc} indicates that  for a single messenger
pair in $5 \oplus \overline{5}$ of SU(5) requires a stop
mass beyond $4$ TeV  to obtain a $123$ GeV Higgs that
tantamount to a considerable fine-tuning.
 We note
that the benchmark points given in Table~\ref{table3} implies an improvement 
in the fine-tuning at $\Delta / \Delta_{mGMSB}\simeq 0.03-0.11.$ Thus an
order of magnitude improvement in fine-tuning can be achieved through
the messenger-matter interactions discussed in this paper.  

Within the framework of GMSB models, the collider constraints on the
sparticle masses are rather model sensitive, crucially depending on
the NLSP \cite{Kats:2011qh}.  Interestingly we observe that among the
preferred models only one has a  nutralino NLSP while the
rest have  slepton NLSPs\footnote{As a small
  digression it is curious to note that a light stau can lead to
  enhanced contribution to the $Br(h\rightarrow \gamma \gamma)$ rate
  through loop contributions in the large $\tan\beta$ regime
  \cite{Carena:2012gp} and can explain the slight excess over SM
  prediction indicated by the present experimental data.}.  The search
strategies and hence the consequent constrains are different for the
two scenarios. In case of a bino NLSP one expects a $2\gamma +
E_T^{miss}$ signal, the limits were discussed in a fairly model
independent manner through \textit{simplified models} in
\cite{Barnard:2012au}. The limits presented for the natural SUSY
(light stop) scenario indicate $m_{\tilde g} > 1.1$ TeV and
$m_{\tilde t_1}> 700$ GeV.  Subsequent updates \cite{:2012afb,:2012mx}
marginally enhance the exclusion on the gluino.  The models with
slepton NLSP are constrained from $SS/OS~ dileptons + E_T^{miss}$ and
$\tau + leptons + jets + E_T^{miss} $ searches. Conservative limits
around $m_{\tilde g} > 800$ GeV were suggested in
\cite{Kats:2011qh}. Recent updates from the CMS, see for example
\cite{Chatrchyan:2012sa,:2012th,Chatrchyan:2012te} and the ATLAS, see
\cite{ATLAS:2012ht} will
certainly increase this bound. However we could not find a
comprehensive model independent study of this scenario in terms of
simplified models in the literature.

In Table~\ref{table3} we present indicative benchmark points for the
four preferred models.  Clearly the limits on the neutralino NLSP
scenario already push the fine-tuning of the models beyond 10. Let us
also point out that recent study on the gravitino phenomenology
indicates that a cosmologically safe upper limit gravitino mass might
be as low  as 16 eV \cite{Viel:2005qj}.  The gravitino mass is given by
the usual relation $m_{3/2} =F/(\lambda_{\phi} \sqrt{3}M_{Pl})$ where
$F$ and $\lambda_{\phi}$ are defined in Eq.~\ref{master_lang} and
$M_{Pl}$ is the Planck mass. Comparing with the gaugino masses given
in Eq.~\ref{usual} we see that,
\begin{equation}
 \frac{m_{3/2}}{M_{\tilde{g}}}\propto
\frac{M}{M_{Pl}}\frac{1}{N_{mess}d},
\end{equation}
where $M$,$N_{mess}$ and $d$ are the messenger scale, messenger multiplicity and the Dynkin
index respectively.  It is clear that the restrictive  limits from
 the gravitino mass can be accommodated in models with higher
messenger multiplicity and/or high Dynkin index. It might be expected
that the large gravitino mass  (e.g. the benchmark point for
Model 5 in Table~\ref{table3}) for models with low Dynkin index can be
handled by considering higher messenger multiplicity. A detailed study
of the effect of varying the number of messengers is beyond the scope
of this paper.

Note that most of the discussion in this paper corresponds to this
light gravitino scenario.  However as the gravitino mass crosses $\sim
50 MeV$ the NLSP becomes increasingly stable leading to stronger
constraints from searches for long living particles at the LHC. For
example a quasi-stable stau has the constraints  $m_{\tilde \tau_1}> 223$ GeV
on its mass \cite{Chatrchyan:2012sp}.  Consequently
this would rule out most of the low fine-tuned regions in our models
providing an added motivation to consider low scale of supersymmetry
breaking.

\section{Conclusion}
Supersymmetric scenarios with viable UV complete supersymmetry
breaking sectors are severely challenged by the measured Higgs mass at
the LHC. Several extension of these models that conjecture the
existence of extra exotic particles have been proposed in the
literature to address this issue
\cite{Moroi:2011aa ,Ellwanger:2012ke,Bae:2012ir,Kyae:2012rv,
Bhattacharyya:2012qj,Endo:2012cc} . Within gauge mediated models one can
evade these constraints economically, by considering messenger-matter
interactions. Some of these models have also been studied in the
recent past ,
\cite{Kang:2012ra,Albaid:2012qk,Han:1998xy,Craig:2012xp,Abdullah:2012tq
  ,Evans:2011bea,Martin:2012dg}.  In the present paper we study a class of
these
models that can be embedded in SU(5) GUT group. We find R-parity and
flavor symmetries provide an organized way to study these models.
Interestingly the flavor symmetry can be tied to the origin of the
observed SM fermion masses and mixing through a Froggatt-Nielsen like
framework. We construct all the possible invariant terms with the
messengers in the 1, 5 and 10 dimensional representations of the SU(5)
group.  Some models or their close variants studied in the literature
show up naturally within this framework while a whole set of new
models are predicted. However it should be emphasized that the choice
of flavor symmetries presented here is not unique. They are presented
mostly in the spirit of proof of principle. It might as well be that
the form of the interaction superpotential discussed here arises from
a completely different underlying flavor structure. Nevertheless the
phenomenological features studied in the second half of the paper are
independent of these assumptions.

A detailed numerical study of the relevant parameter space of the
different scenarios is carried out to compare these models. We identify models
that can effectively raise the Higgs mass to the favoured range without
admitting too much fine-tuning. We find many of the generic and
specific SUSY searches at LHC imply considerable constraints on these
models.  Incidentally we observe that all but one of the preferred
models have a slepton NLSP.  However we could not locate a detailed
model independent study of the LHC constraints on this scenario in the
literature. Most studies are restricted to benchmark points or related
to details of the entire spectrum.  Considering that many other
possible \textit{natural} SUSY models within the context of GMSB could
lead to a slepton NLSP, a detailed study of this scenario is highly
anticipated.

\noindent {\bf Acknowledgments:}~ We thank St\'ephane Lavignac, 
and Gautam Bhattacharyya  for reading the manuscript and
suggesting valuable changes at different stages of the work. We also thank James
Barnard and Kamakhya Prasad Modok for useful discussions. TSR 
acknowledges hospitality at Department of Physics, Calcutta
University, during the early stages of the work.  The research of
TSR is supported the Australian Research Council.


\vspace{2cm}
 \begin{thebibliography}{99}

\bibitem{:2012gk}
  G.~Aad {\it et al.}  [ATLAS Collaboration],
  ``Observation of a new particle in the search for the Standard Model Higgs
boson with the ATLAS detector at the LHC,''
  Phys.\ Lett.\ B {\bf 716} (2012) 1
  [arXiv:1207.7214 [hep-ex]].

\bibitem{:2012gu}
  S.~Chatrchyan {\it et al.}  [CMS Collaboration],
  ``Observation of a new boson at a mass of 125 GeV with the CMS experiment at
the LHC,''
  Phys.\ Lett.\ B {\bf 716} (2012) 30
  [arXiv:1207.7235 [hep-ex]].
  
 
 

\bibitem{Draper:2011aa}
  P.~Draper, P.~Meade, M.~Reece and D.~Shih,
  ``Implications of a 125 GeV Higgs for the MSSM and Low-Scale SUSY Breaking,''
  Phys.\ Rev.\ D {\bf 85} (2012) 095007
  [arXiv:1112.3068 [hep-ph]].
  

 %4
\bibitem{Giudice:1998bp} 
  G.~F.~Giudice and R.~Rattazzi,
  ``Theories with gauge mediated supersymmetry breaking,''
  Phys.\ Rept.\  {\bf 322}, 419 (1999)
  [hep-ph/9801271].
  
  
\bibitem{Intriligator:2010be}
  K.~Intriligator and M.~Sudano,
  ``General Gauge Mediation with Gauge Messengers,''
  JHEP {\bf 1006} (2010) 047
  [arXiv:1001.5443 [hep-ph]].
  
  
  
\bibitem{Bajc:2012sm}
  B.~Bajc, S.~Lavignac and T.~Mede,
  ``Supersymmetry Breaking Induced by Radiative Corrections,''
  JHEP {\bf 1207} (2012) 185
  [arXiv:1202.2845 [hep-ph]].
  
 

\bibitem{Barbieri:1995tw}
  R.~Barbieri, L.~J.~Hall and A.~Strumia,
  ``Violations of lepton flavor and CP in supersymmetric unified theories,''
  Nucl.\ Phys.\ B {\bf 445} (1995) 219
  [hep-ph/9501334],
 
\bibitem{Barbieri:1995rs}
  R.~Barbieri, L.~J.~Hall and A.~Strumia,
  ``Hadronic flavor and CP violating signals of superunification,''
  Nucl.\ Phys.\ B {\bf 449} (1995) 437
  [hep-ph/9504373],
  
\bibitem{ArkaniHamed:1995fs}
  N.~Arkani-Hamed, H.~-C.~Cheng and L.~J.~Hall,
 ``Flavor mixing signals for realistic supersymmetric unification,''
  Phys.\ Rev.\ D {\bf 53} (1996) 413
  [hep-ph/9508288].
 
 
  
 
\bibitem{Khlopov:1984pf}
  M.~Y.~.Khlopov and A.~D.~Linde,
  ``Is It Easy to Save the Gravitino?,''
  Phys.\ Lett.\ B {\bf 138} (1984) 265.
  

  
 %6 
\bibitem{Kang:2012ra} 
  Z.~Kang, T.~Li, T.~Liu, C.~Tong and J.~M.~Yang,
  ``A Heavy SM-like Higgs and a Light Stop from Yukawa-Deflected Gauge
Mediation,''
  Phys.\ Rev.\ D {\bf 86}, 095020 (2012)
  [arXiv:1203.2336 [hep-ph]].

\bibitem{Albaid:2012qk} 
  A.~Albaid and K.~S.~Babu,
  ``Higgs boson of mass 125 GeV in GMSB models with messenger-matter mixing,''
  arXiv:1207.1014 [hep-ph].

\bibitem{Han:1998xy} 
  T.~Han and R.~-J.~Zhang,
  ``Direct messenger - matter interactions in gauge - mediated supersymmetry
breaking models,''
  Phys.\ Lett.\ B {\bf 428}, 120 (1998)
  [hep-ph/9802422].

\bibitem{Craig:2012xp} 
  N.~Craig, S.~Knapen, D.~Shih and Y.~Zhao,
  ``A Complete Model of Low-Scale Gauge Mediation,''
  arXiv:1206.4086 [hep-ph].


\bibitem{Abdullah:2012tq} 
  M.~Abdullah, I.~Galon, Y.~Shadmi and Y.~Shirman,
  ``Flavored Gauge Mediation, A Heavy Higgs, and Supersymmetric Alignment,''
  arXiv:1209.4904 [hep-ph].

\bibitem{Evans:2011bea} 
  J.~L.~Evans, M.~Ibe and T.~T.~Yanagida,
  ``Relatively Heavy Higgs Boson in More Generic Gauge Mediation,''
  Phys.\ Lett.\ B {\bf 705}, 342 (2011)
  [arXiv:1107.3006 [hep-ph]]. 

\bibitem{Martin:2012dg} 
  S.~P.~Martin and J.~D.~Wells,
  ``Implications of gauge-mediated supersymmetry breaking with vector-like
quarks and a ~125 GeV Higgs boson,''
  Phys.\ Rev.\ D {\bf 86}, 035017 (2012)
  [arXiv:1206.2956 [hep-ph]].

  
  %\cite{Picariello:1998dy}
\bibitem{Picariello:1998dy}
  M.~Picariello and A.~Strumia,
  ``Next-to-leading order corrections to gauge mediated gaugino masses,''
  Nucl.\ Phys.\ B {\bf 529} (1998) 81
  [hep-ph/9802446].
  %%CITATION = HEP-PH/9802446;%%

  %\cite{Lee:2011dw}
\bibitem{Lee:2011dw}
  J.~Y.~Lee and Y.~W.~Yoon,
  ``Next-to-leading order contributions to the pole mass of gluino in minimal gauge mediation,''
  JHEP {\bf 1205} (2012) 029
  [arXiv:1112.3904 [hep-ph]].
  %%CITATION = ARXIV:1112.3904;%%

  
 
  
\bibitem{Calibbi:2009cp} 
  L.~Calibbi, L.~Ferretti, A.~Romanino and R.~Ziegler,
  ``Gauge coupling unification, the GUT scale, and magic fields,''
  Phys.\ Lett.\ B {\bf 672}, 152 (2009)
  [arXiv:0812.0342 [hep-ph]].

\bibitem{Byakti:2012qk} 
  P.~Byakti and D.~Ghosh,
  ``Magic Messengers in Gauge Mediation and signal for 125 GeV boosted Higgs
boson,''
  Phys.\ Rev.\ D {\bf 86}, 095027 (2012)
  [arXiv:1204.0415 [hep-ph]].
  
 
\bibitem{Froggatt:1978nt}
  C.~D.~Froggatt and H.~B.~Nielsen,
 ``Hierarchy of Quark Masses, Cabibbo Angles and CP Violation,''
  Nucl.\ Phys.\ B {\bf 147} (1979) 277.
 
\bibitem{Savoy:2010sj}
  C.~A.~Savoy and M.~Thormeier,
  ``Exotic particles below the TeV from low scale flavour theories,''
  JHEP {\bf 1103} (2011) 128
  [arXiv:1003.2090 [hep-ph]],
  
\bibitem{Eboli:2011hr}
  O.~J.~P.~Eboli, C.~A.~Savoy and R.~Zukanovich Funchal,
  ``A Rationale for Long-lived Quarks and Leptons at the LHC: Low Energy Flavour
Theory,''
  JHEP {\bf 1202} (2012) 123
  [arXiv:1112.5108 [hep-ph]],
  
\bibitem{Ray:2011dm}
  T.~S.~Ray, H.~de Sandes and C.~A.~Savoy,
  ``Gluino, Wino and Higgsino-Like Particles without Supersymmetry,''
  Phys.\ Lett.\ B {\bf 712} (2012) 401
  [arXiv:1112.6180 [hep-ph]].
 

\bibitem{Dimopoulos:1996ig}
  S.~Dimopoulos and G.~F.~Giudice,
 ``Multimessenger theories of gauge mediated supersymmetry breaking,''
  Phys.\ Lett.\ B {\bf 393} (1997) 72
  [hep-ph/9609344].
   

 \bibitem{Giudice:1997ni} 
  G.~F.~Giudice and R.~Rattazzi,
  ``Extracting supersymmetry breaking effects from wave function
renormalization,''
  Nucl.\ Phys.\ B {\bf 511}, 25 (1998)
  [hep-ph/9706540].

  \bibitem{Chacko:2001km} 
  Z.~Chacko and E.~Ponton,
  ``Yukawa deflected gauge mediation,''
  Phys.\ Rev.\ D {\bf 66}, 095004 (2002)
  [hep-ph/0112190].


  %\cite{Evans:2013kxa}
\bibitem{Evans:2013kxa}
  J.~A.~Evans and D.~Shih,
  ``Surveying Extended GMSB Models with mh=125 GeV,''
  arXiv:1303.0228 [hep-ph].
  %%CITATION = ARXIV:1303.0228;%%
  %2 citations counted in INSPIRE as of 12 Apr 2013


\bibitem{Evans:2012hg}
  J.~L.~Evans, M.~Ibe, S.~Shirai and T.~T.~Yanagida,
  ``A 125GeV Higgs Boson and Muon g-2 in More Generic Gauge Mediation,''
  Phys.\ Rev.\ D {\bf 85} (2012) 095004
  [arXiv:1201.2611 [hep-ph]]. 

\bibitem{Mohapatra:1998rq} 
  R.~N.~Mohapatra and P.~B.~Pal,
  ``Massive neutrinos in physics and astrophysics. Second edition,''
  World Sci.\ Lect.\ Notes Phys.\  {\bf 60}, 1 (1998)
  [World Sci.\ Lect.\ Notes Phys.\  {\bf 72}, 1 (2004)].
 
 
 %\cite{Joaquim:2006uz}
\bibitem{Joaquim:2006uz}
  F.~R.~Joaquim and A.~Rossi,
  ``Gauge and Yukawa mediated supersymmetry breaking in the triplet seesaw scenario,''
  Phys.\ Rev.\ Lett.\  {\bf 97} (2006) 181801
  [hep-ph/0604083].
  %%CITATION = HEP-PH/0604083;%%
  
  %\cite{Joaquim:2006mn}
\bibitem{Joaquim:2006mn}
  F.~R.~Joaquim and A.~Rossi,
  ``Phenomenology of the triplet seesaw mechanism with Gauge and Yukawa mediation of SUSY breaking,''
  Nucl.\ Phys.\ B {\bf 765} (2007) 71
  [hep-ph/0607298].
  %%CITATION = HEP-PH/0607298;%%

 

\bibitem{Martin:1997ns}
  S.~P.~Martin,
  ``A Supersymmetry primer,''
  In *Kane, G.L. (ed.): Perspectives on supersymmetry II* 1-153
  [hep-ph/9709356].
  

\bibitem{Delgado:2007rz}
  A.~Delgado, G.~F.~Giudice and P.~Slavich,
  ``Dynamical mu term in gauge mediation,''
  Phys.\ Lett.\ B {\bf 653} (2007) 424
  [arXiv:0706.3873 [hep-ph]].
  
\bibitem{Djouadi:2002ze}
  A.~Djouadi, J.~-L.~Kneur and G.~Moultaka,
  ``SuSpect: A Fortran code for the supersymmetric and Higgs particle spectrum
in the MSSM,''
  Comput.\ Phys.\ Commun.\  {\bf 176} (2007) 426
  [hep-ph/0211331].
  
  

\bibitem{Belanger:2010gh}
  G.~Belanger, F.~Boudjema, P.~Brun, A.~Pukhov, S.~Rosier-Lees, P.~Salati and A.~Semenov,
  ``Indirect search for dark matter with micrOMEGAs2.4,''
  Comput.\ Phys.\ Commun.\  {\bf 182} (2011) 842
  [arXiv:1004.1092 [hep-ph]].
  
\bibitem{Belanger:2004yn}
  G.~Belanger, F.~Boudjema, A.~Pukhov and A.~Semenov,
 ``micrOMEGAs: Version 1.3,''
  Comput.\ Phys.\ Commun.\  {\bf 174} (2006) 577
  [hep-ph/0405253].
   
\bibitem{Barbieri:1987fn}
  R.~Barbieri and G.~F.~Giudice,
  ``Upper Bounds on Supersymmetric Particle Masses,''
  Nucl.\ Phys.\ B {\bf 306} (1988) 63.
  
 
\bibitem{Wright:1998mk}
  D.~Wright,
  ``Naturally nonminimal supersymmetry,''
  hep-ph/9801449.
  
  
\bibitem{Ajaib:2012vc}
  M.~A.~Ajaib, I.~Gogoladze, F.~Nasir and Q.~Shafi,
  ``Revisiting mGMSB in Light of a 125 GeV Higgs,''
  Phys.\ Lett.\ B {\bf 713} (2012) 462
  [arXiv:1204.2856 [hep-ph]].
   
  
\bibitem{Kats:2011qh}
  Y.~Kats, P.~Meade, M.~Reece and D.~Shih,
  ``The Status of GMSB After 1/fb at the LHC,''
  JHEP {\bf 1202} (2012) 115
  [arXiv:1110.6444 [hep-ph]].
    
  \bibitem{Carena:2012gp}
  M.~Carena, S.~Gori, N.~R.~Shah, C.~E.~M.~Wagner and L.~-T.~Wang,
  ``Light Stau Phenomenology and the Higgs $\gamma\gamma$ Rate,''
  JHEP {\bf 1207} (2012) 175
  [arXiv:1205.5842 [hep-ph]].
  
 
\bibitem{Barnard:2012au}
  J.~Barnard, B.~Farmer, T.~Gherghetta and M.~White,
  ``Natural gauge mediation with a bino NLSP at the LHC,''
  arXiv:1208.6062 [hep-ph].
   

\bibitem{:2012afb}
  G.~Aad {\it et al.}  [ATLAS Collaboration],
 ``Search for diphoton events with large missing transverse momentum in 7 TeV
proton-proton collision data with the ATLAS detector,''
  arXiv:1209.0753 [hep-ex],
 
\bibitem{:2012mx}
  S.~Chatrchyan {\it et al.}  [CMS Collaboration],
  ``Search for new physics in events with photons, jets, and missing transverse
energy in pp collisions at sqrt(s) = 7 TeV,''
  arXiv:1211.4784 [hep-ex].
 
 
\bibitem{Chatrchyan:2012sa}
  S.~Chatrchyan {\it et al.}  [CMS Collaboration],
  ``Search for new physics in events with same-sign dileptons and $b$-tagged
jets in $pp$ collisions at $\sqrt{s}=7$ TeV,''
  JHEP {\bf 1208} (2012) 110
  [arXiv:1205.3933 [hep-ex]],
 
\bibitem{:2012th}
  S.~Chatrchyan {\it et al.}  [CMS Collaboration],
  ``Search for new physics with same-sign isolated dilepton events with jets and
missing transverse energy,''
  Phys.\ Rev.\ Lett.\  {\bf 109} (2012) 071803
  [arXiv:1205.6615 [hep-ex]],
 
\bibitem{Chatrchyan:2012te}
  S.~Chatrchyan {\it et al.}  [CMS Collaboration],
  ``Search for new physics in events with opposite-sign leptons, jets, and
missing transverse energy in pp collisions at sqrt(s) = 7 TeV,''
  arXiv:1206.3949 [hep-ex].
 
 
\bibitem{ATLAS:2012ht}
  G.~Aad {\it et al.}  [ATLAS Collaboration],
  ``Search for Supersymmetry in Events with Large Missing Transverse Momentum,
Jets, and at Least One Tau Lepton in 7 TeV Proton-Proton Collision Data with the
ATLAS Detector,''
  arXiv:1210.1314 [hep-ex].
  

\bibitem{Chatrchyan:2012sp}
  S.~Chatrchyan {\it et al.}  [CMS Collaboration],
 ``Search for heavy long-lived charged particles in pp collisions at sqrt(s)=7
TeV,''
  Phys.\ Lett.\ B {\bf 713} (2012) 408
  [arXiv:1205.0272 [hep-ex]].
 
\bibitem{Viel:2005qj}
  M.~Viel, J.~Lesgourgues, M.~G.~Haehnelt, S.~Matarrese and A.~Riotto,
  ``Constraining warm dark matter candidates including sterile neutrinos and
light gravitinos with WMAP and the Lyman-alpha forest,''
  Phys.\ Rev.\ D {\bf 71} (2005) 063534
  [astro-ph/0501562].
 %I am here.  
  
  
  
\bibitem{Moroi:2011aa}
  T.~Moroi, R.~Sato and T.~T.~Yanagida,
  ``Extra Matters Decree the Relatively Heavy Higgs of Mass about 125 GeV in the
Supersymmetric Model,''
  Phys.\ Lett.\ B {\bf 709} (2012) 218
  [arXiv:1112.3142 [hep-ph]].
  
\bibitem{Ellwanger:2012ke}
  U.~Ellwanger and C.~Hugonie,
  ``Higgs bosons near 125 GeV in the NMSSM with constraints at the GUT scale,''
  Adv.\ High Energy Phys.\  {\bf 2012} (2012) 625389
  [arXiv:1203.5048 [hep-ph]].
  
\bibitem{Bae:2012ir}
  K.~J.~Bae, T.~H.~Jung and H.~D.~Kim,
  ``125 GeV Higgs as a pseudo-Goldstone boson in supersymmetry with vector-like
matters,''
  arXiv:1208.3748 [hep-ph].
  
\bibitem{Kyae:2012rv}
  B.~Kyae and J.~-C.~Park,
 ``A Singlet-Extension of the MSSM for 125 GeV Higgs with the Least Tuning,''
  arXiv:1207.3126 [hep-ph].
  
\bibitem{Bhattacharyya:2012qj}
  G.~Bhattacharyya and T.~S.~Ray,
  ``Pushing the SUSY Higgs mass towards 125 GeV with a color adjoint,''
  arXiv:1210.0594 [hep-ph],
  
\bibitem{Endo:2012cc}
  M.~Endo, K.~Hamaguchi, K.~Ishikawa, S.~Iwamoto and N.~Yokozaki,
  ``Gauge Mediation Models with Vectorlike Matters at the LHC,''
  arXiv:1212.3935 [hep-ph].
    
  
\end{thebibliography}
\end{document}